\documentclass[smallextended]{svjour3}
\smartqed
\usepackage{graphicx}
\usepackage{amsmath,amssymb}
\usepackage{longtable}
\usepackage{booktabs}
\usepackage{comment}
\usepackage{array}
\usepackage{enumitem}
\usepackage{url}

\newcommand{\code}[1]{\texttt{#1}}
\newcommand{\diag}{\mathop{\mathrm{diag}}\nolimits}
\newcolumntype{L}[1]{>{\raggedright\arraybackslash}p{#1}}
\newcolumntype{R}[1]{>{\raggedleft\arraybackslash}p{#1}}

\makeatletter
\renewcommand{\makeheadbox}{}
\makeatother

\begin{document}

\title{Quadratic unconstrained binary optimization benchmark problems from mixed Hamming packings}
\titlerunning{QUBO benchmarks from mixed Hamming packing problems}
\authorrunning{P. Naszvadi et al.}

\author{P\'{e}ter Naszvadi \and
        Marcell Tukora \and
        Roman Ku\v{z}el \and
        M\'{a}ty\'{a}s Koniorczyk}

\institute{P. Naszvadi \at
               HUN-REN Wigner Research Centre for Physics, Budapest, Hungary \\
               Faculty of Informatics, E\"{o}tv\"{o}s Lor\'{a}nd University, Budapest, Hungary \\
               \email{naszvadi.peter@wigner.hun-ren.hu}
           \and
           M. Tukora \at
               HUN-REN Wigner Research Centre for Physics, Budapest, Hungary \\
               Corvinus University of Budapest, Budapest, Hungary \\
               \email{tukora.marcell@wigner.hun-ren.hu}
           \and
           M. Koniorczyk \at
               HUN-REN Wigner Research Centre for Physics, Budapest, Hungary \\
               \email{koniorczyk.matyas@wigner.hun-ren.hu}
           \and
           R. Ku\v{z}el \at
               Centre for Applied Quantum Computing, Rudolfovo -- Science and Technology Centre, Novo Mesto, Slovenia \\
               \email{roman.kuzel@rudolfovo.eu}
}

\date{20 November 2026}

\maketitle

\begin{abstract}
  Benchmarking exact solvers and noisy intermediate-scale quantum
  devices for quadratic unconstrained binary optimization (QUBO)
  requires small yet challenging instances, preferably coming from
  actual applications. We propose such a benchmark set derived from
  mixed Hamming packing, a hard problem in coding theory. Starting
  from its mixed-integer linear programming formulation, we provide a
  generator for the equivalent QUBO models and study 133 instances
  whose optima are certified by that formulation. In the feature space
  of the MQLib library, we position a broad range of existing
  quantum-oriented benchmarks relative to the classical ones: they lie
  along a narrow, sparse border of the region covered by MQLib,
  whereas our instances form a coherent family in a region that both
  collections leave almost empty. Classical heuristics and a simulated
  bifurcation machine find the optimum of most instances quickly, in
  several cases sooner than an exact mixed-integer solver first
  encounters it, suggesting heuristic warm starts for exact solvers;
  for an exact QUBO solver, in turn, it appears challenging to prove
  optimality. In Ising form, the instances are densely connected
  antiferromagnets whose large local fields are almost exactly
  balanced by the couplings, so that the decision on each variable
  rests on a small remainder. This makes them hard to embed and to
  represent accurately on current quantum annealers, in line with our
  D-Wave experiments.  
\end{abstract}

\section{Introduction}

Quadratic unconstrained binary optimization (QUBO) problems,
\begin{equation}
  \label{eq:QUBO}
  \min_{\mathbf{x}} \mathbf{x}^\top Q\,\mathbf{x},\quad \mathbf{x}\in \{0,1\}^{N},
\end{equation}
have an extensive literature in operations research, see~\cite{Punnen}
for a comprehensive review. They are NP-hard in the worst case, and
they can be considered as one of the universal encodings for
combinatorial problems in NP. They are equivalent to Ising problems
\begin{equation}
    \label{eq:Ising}
    \min_{\mathbf{s}} \; \mathbf{s}^\top J \mathbf{s} + \mathbf{h}^\top \mathbf{s} + c,\quad \mathbf{s}\in \{-1,+1\}^{N},
\end{equation}
as well as to Max-Cut problems. In recent decades, many heuristic
algorithms have been developed to solve such problems. Dunning et
al.~\cite{Dunning2018} provide a comprehensive collection of such
heuristic algorithms, called MQLib, supplemented by an efficient
open-source C++ implementation designed with the needs of benchmarking
experiments in mind.

There have been a number of QUBO benchmarks proposed for heuristic
(and exact) QUBO solvers, G-set~\cite{Helmberg2000} being probably the
most widely used one. QPLIB~\cite{Furini2019} is another widely used
library, covering general quadratic programming instances including
unconstrained binary ones. The work of Dunning et
al.~\cite{Dunning2018} also provides an extensive and systematic
survey of QUBO benchmark sets. They also introduce a feature set based
on graph-theoretic features of the Max-Cut equivalent of the problem,
which can be helpful in categorizing features of different benchmarks.

Besides heuristic approaches, exact QUBO solvers have also seen
relevant development recently. Commercial solvers such as
Gurobi~\cite{Gurobi} have remarkable capabilities, and
BiqBin~\cite{Gusmeroli2022}, a highly parallel solver aimed at
high-performance computing environments, is in active
development~\cite{BiqBinRepo}. Testing and benchmarking these solvers
requires small-sized yet challenging QUBO instances.

With the advent of quantum computing, the relevance of QUBO problems
and their solvers has gained further research attention, as this is
the very problem quantum annealers and certain gate-based algorithms
such as QAOA~\cite{Farhi2014} can solve. In the present era of noisy
intermediate-scale quantum (NISQ) devices, especially the simple
usability of D-Wave quantum annealers has generated a large number of
publications that address the encoding of various problems in QUBO
form and explore the possibility of solving these with current quantum
hardware.

Concerning benchmarks, NISQ quantum devices need problems different
from those designed for classical solvers. First, as far as NISQ
devices are considered, the benchmark problems should be small, that
is, a few hundred variables. In addition, they should pose a challenge
to quantum hardware. Tile-planting
instances~\cite{Perera2020,Hamze2018} are square- or cubic-lattice
Ising spin glasses assembled from tiles that share a common ground
state, so the planted optimum is known by construction while the
fraction of frustrated tiles can be tuned continuously, pushing the
problem through an easy-to-hard transition and thereby challenging
quantum annealers with a genuinely rugged energy landscape independent
of instance size. Higher-order binary optimization instances built
directly on the heavy-hex qubit topology of superconducting
processors~\cite{Chamberland2020} match the native two- and
three-qubit interaction sets of such gate-based hardware, avoiding
routing overhead and thus probing the hardware's capability directly,
a construction recently used to substantiate claims of runtime quantum
advantage for digital quantum optimization~\cite{Chandarana2025}.  One
should remark here that the benchmarking methodologies of quantum
devices are still in development~\cite{Tuziemski2025}. It is a common
feature of quantum-aimed benchmarks that they do not come from actual
independent applications but are designed on purpose to be both
suitable for, and interesting from the point of view of, the actual
quantum computing setup.

Recently, a number of new physics-motivated heuristics have been
proposed that show quantum-like behavior from certain points of
view. Among these, the simulated bifurcation machine (SBM) is one of
the most efficient
ones~\cite{Goto_2019,SBM,Kanao_2022,Pawlowski2026}. Its idea is rooted
in a proposal for a quantum annealing implementation on coupled
nonlinear oscillators~\cite{Goto_2016}. This is a quantum system that
can be understood as a canonically quantized version of a classical
mechanical system. The differential equation describing its time
evolution can be solved with the symplectic Euler method. It is
possible to efficiently implement this method on GPUs or FPGAs. The
mechanical system itself exhibits bifurcations, thereby realizing a
minimum search different from those in classical heuristics. One can
depart from this physical picture to obtain more efficient heuristics:
a discrete and a ballistic version also exist~\cite{SBM}. These
heuristics show excellent performance, and are capable of addressing
large and dense QUBO instances without the need for
embedding. Recently, further extensions have been proposed to make
simulated bifurcation more
``quantum-like''~\cite{Kanao_2022,Pawlowski2026}.

The motivation of the present work had originally been to propose a
benchmark set which contains small and challenging QUBO instances
coming from an application. Such problems can then be interesting from
the point of view of NISQ quantum device benchmarking as well as from
the point of view of exact QUBO solvers. To warrant the challenging
nature of our problem instances, we were looking for a mathematical
problem whose instances are known to be hard to solve---we opted for
the so-called mixed Hamming packings.

QUBO encodings of code-theoretic search problems have been considered
before. Most closely related to our work, Yukiyoshi and
Ishikawa~\cite{yukiyoshi2022quantumsearchalgorithmbinary} formulated
the search for a certain class of binary error-correcting codes as a
QUBO problem, with the aim of solving it with Grover adaptive search
on a future fault-tolerant quantum computer, and demonstrated the
approach in classical simulation on a single small example. Our
motivation is different: rather than a quantum algorithm for a single
problem, we seek a family of instances that can serve as a benchmark
for currently available quantum and classical QUBO solvers.

In a previous paper~\cite{Naszvadi2025} we described our approach to
the problem: we extended a simple linear 0-1 programming formulation,
which is straightforward but known to be very inefficient, with
additional constraints that make it suitable for addressing smaller
but already nontrivial instances. We also introduced a decomposition
method that enables bigger problem instances to be subdivided into
smaller ones.

Importantly, our linear 0-1 problem has a QUBO equivalent. The present
contribution analyzes this formulation, considering the mixed integer
programming (MILP) formulation as a baseline. In this way we propose a
QUBO benchmark set which encodes instances of a relevant mathematical
problem, whose smaller instances, as will be demonstrated here, can be
of interest as small yet challenging benchmarks.

In order to compare our instances with others, motivated by the
results of Dunning et al.~\cite{Dunning2018}, we have conducted an
analysis to place our benchmark set, and other quantum-aimed
benchmarks, on the `map' of known benchmark problems in the feature
space of properties relevant from the classical heuristics' point of
view. This analysis shows that our instances form a coherent family
that is markedly different from the other benchmarks.

In the light of the availability of the comprehensive collection of
classical QUBO heuristics, as well as the recent development of
other physics-motivated solvers, we also found it important to address
our problems with classical heuristics as well as with a simulated
bifurcation machine. 

This paper is organized as follows. In
Section~\ref{sec:codetheoreticproblem} we introduce a code-theoretic
problem together with its state-of-the-art modeling techniques. In
Section~\ref{sec:hammingqubobenchmark} our benchmark set is introduced
and compared to others. Section~\ref{sec:benchmarking} contains
numerical results ranging from the MILP baseline through an exact QUBO
solver (Gurobi), a wide range of classical heuristics and a simulated
bifurcation machine, to D-Wave quantum annealers. Finally, in
Section~\ref{sec:conclusions} the results are summarized, conclusions
are drawn, and further research directions are outlined. A worked
example of the model generator and the complete list of instances are
given in Online Resource~1. The data and codes to reproduce the
results, including the actual benchmark instances and the scripts to
generate them, are available online; see the "data availability"
section.

\section{A code-theoretic problem and its MILP model}
\label{sec:codetheoreticproblem}

Here we describe a problem in coding theory, from which our QUBO and
MILP models originate, and briefly recall the specific reduction
technique and results of our previous paper~\cite{Naszvadi2025}, on
which the model generator described in
Section~\ref{sec:hammingqubobenchmark} is built.

\subsection{Problem statement}

We consider $k$-ary alphabets $\mathbb{Z}_k=\{0,1,\ldots,k-1\}$ without
assuming any algebraic structure, following the notation of
\cite{Naszvadi2025}.

\begin{definition}\emph{Hamming space.}\label{def:hammingspace}
  Given cardinalities $2\le k_1, k_2,\dots, k_n<\infty$, the Hamming
  space
  $H=H(k_1,k_2,\ldots,k_n):=\mathbb{Z}_{k_1}\mathbb{Z}_{k_2}\cdots\mathbb{Z}_{k_n}$
  is the Cartesian product of the corresponding alphabets. Its
  elements are called codewords or words. 
\end{definition}
We will follow the convention that $k_1\geq k_2\ldots\geq k_n$
throughout this paper which can be done w.l.o.g.
\begin{definition}\emph{Hamming distance.}
For $w,w'\in H$,
\begin{equation}
    d_{H}(w,w')=\sum\limits_{
    \substack{j=1 \\ w[j]\neq w'[j]}}^{n} 1
\end{equation}
is the number of positions at which $w$ and $w'$ differ.
\end{definition}

A nonempty subset $C\subseteq H$ is a code; if $|C|\ge2$, its minimum
distance is defined as $d_{H}(C)=\min_{w\neq w'\in C}d_{H}(w,w')$.

\begin{definition}\emph{Hamming ball, (closed) ball or Hamming-sphere.}
For $p\in H$ and an integer $r\ge 0$,
\begin{equation} \label{eq:hammingball}
    B(p,r)=\{w\in H : d_H(w,p)\leq r\}.
\end{equation}
\end{definition}

\begin{definition}\emph{Optimal Hamming packing.}
\begin{equation} \label{eq:hammingpackingproblem}
    N(H;d) = \max\limits_{d_{H}(C)\geq d}{|C|}.
\end{equation}
\end{definition}

Determining $N(H;d)$ for a given $H$ and $d\in\mathbb{N}:1\le d\le n$
is a key question in coding theory, studied extensively for uniform
alphabets~\cite{Best1978,Bogdanova2001,Gijswijt2006,Laaksonen2017} and
also for the mixed case we are considering
here~\cite{Brouwer1998,Oestergaard2000,Perkins2006,Litjens2018,BrouwerTable,Naszvadi2025,Oestergaard2009,Oestergaard2010}.

Our starting point here is the textbook MILP formulation of
  Eq.~\eqref{eq:hammingpackingproblem}~\cite{Oestergaard2005,Naszvadi2025}.
  Introducing a binary indicator $x_a$ for every $a\in H$, with $x_a=1$ iff
  $a\in C$, the maximum packing is obtained as any optimal solution of
  \begin{eqnarray}
      \label{eq:naiveilp}
      \max & \sum\limits_{a\in H}{x_a}& \\
      s.t. ~\hfill~ x_a + x_b &\leq 1 & {\forall a,b\in H: }\ {1\leq d_{H}(a,b)\leq d-1} \nonumber \\
      & x\in{\left\{0,1\right\}}^{|H|}& \nonumber.
  \end{eqnarray}
  The objective counts the selected codewords. Two different codewords
  \emph{conflict} if their distance is below $d$. The pairwise mutex
  constraints forbid selecting conflicting word pairs. Therefore,
  \eqref{eq:naiveilp} is the MILP form of the maximum independent set
  problem of the conflict graph, and $N(H;d)$ its independence
  number. It has $|H|=\prod_i k_i$ variables and one constraint per
  edge of the conflict graph, and it is monotone: setting any $x_a=1$
  to $0$ in a feasible solution preserves feasibility. Due to the
  exponential growth of $|H|$ and the large symmetry group of $H$,
  \eqref{eq:naiveilp} is inefficient to solve
  directly~\cite{Naszvadi2025}; the model below extends it with the
  reductions presented in Section~\ref{sec:contactgraphs} without
  changing the optimum.
  
\subsection{Contact graphs and the reduction technique}
\label{sec:contactgraphs}

Though the model in~\eqref{eq:naiveilp} is highly inefficient in
comparison with other specialized methods exploiting the symmetries of
$H$, it is possible to improve it by adding further
constraints. In~\cite{Naszvadi2025} we adopted the following concept
from continuous sphere-packing problems to the present discrete
context.
\begin{definition}\emph{Contact graph.}
Given a code $C$ with minimum distance $d_H(C)=d$, its contact graph
$CG(C)$ has vertex set $C$, with an edge between $v\ne w$ iff
$d_H(v,w)=d$.
\end{definition}
One of the main results of \cite{Naszvadi2025} is that every (not
necessarily optimal) packing can be transformed, without changing its
cardinality, into one whose contact graph is connected.  This property
gives rise to a family of valid constraints that can be added
independently to the MILP without excluding any optimum.  The first
one is a cut constraint forbidding isolated contact-graph vertices
(cf.\ Corollary~4 in \cite{Naszvadi2025}):

\begin{equation}\label{eq:noisolated}
  x_w \;\le\!\! \sum_{d_H(w,w')=d}\!\! x_{w'} \qquad \forall w\in H.
  \end{equation}
  
  The same connectivity property also yields a decomposition strategy:
  once a codeword $p$ is known to belong to some optimal packing, the
  ball $B(p,d-1)$ around $p$, apart from $p$ itself, can be discarded from $H$ (its
  points are at distance $<d$ from $p$, hence excluded from any code
  containing $p$ anyway) while fixing $x_p=1$; branching over small
  connected contact-graph subgraphs and iterating this over a growing
  set of fixed centers can split large instances into smaller,
  independently solvable ones. In the present work we do not apply
  such decompositions as we wanted to keep the sizes of the instances
  under control.
  
  Suppose $H$ is a Hamming space and $d\in\mathbb{N}^+$ a minimal
  distance.  Let $P\subseteq H$ denote the (possibly empty)
  \emph{prune list}: the set of codewords already fixed to belong to
  the packing, as motivated above; in the instances studied in the
  present work it will always contain the all-0 codeword only.  The
  reference MILP model used throughout this paper reads
\begin{eqnarray}
    \label{eq:milpmodel1}
    \max & \sum\limits_{a\in H}{x_a}& \\
    s.t. ~\hfill~  x_p &= 1 &{\forall p \in P}\nonumber \\
    ~  x_a &= 0 & 1\leq d_{H}(a,p) \leq d-1 \ \text{for some } p \in P
    \nonumber \\
    ~ x_a + x_b &\leq 1 & {\forall a,b\in H: }\ {1\leq d_{H}(a,b)\leq d-1} \nonumber \\
      \sum\limits_{d_{H}(a,w)=d}{x_{w}}&\geq x_{a}&\forall a\in H \nonumber \\
    & x\in{\left\{0,1\right\}}^{|H|}& \nonumber,
\end{eqnarray}
whose four constraint blocks are, respectively: fixing the pruned
centers to the packing; excluding the (now redundant) ball of radius
$d-1$ around each of them; the pairwise mutex constraints of the
underlying independent set problem; and the no-isolated-vertex cut
\eqref{eq:noisolated}.

In \cite{Naszvadi2025} we found that this approach, along with other
possible model extensions not employed in the present contribution,
made our model competitive with the specialized algorithms in the case
of smaller problem instances. While the MILP version of our model can
be solved and thus serves as a baseline, it is easy to rewrite into a
QUBO form.

\subsection{QUBO formulation}
\label{sec:optmodels}

Following the standard penalty-based reformulation of pairwise 0--1
exclusion constraints (cf.\ the generic QUBO and Ising forms
\eqref{eq:QUBO}--\eqref{eq:Ising} of Section~1), the mutex constraints of
\eqref{eq:milpmodel1} can be replaced with a quadratic penalty term, giving
an equivalent QUBO model for $N(H;d)$ on the pruned space
$H\setminus\bigcup_{p\in P}B(p,d-1)$:

\begin{equation}\label{eq:qubomodel1}
\min_{x\in\{0,1\}^{|H|}}\ \sum_{a\in H} Q_{aa}\,x_a
+ \sum_{\substack{a,b\in H\\ a<b}} Q_{ab}\,x_a x_b,
\end{equation}
\begin{equation}\label{eq:quboQ}
Q_{aa}=-1,\qquad
Q_{ab}=\begin{cases} \lambda & 1\le d_H(a,b)\le d-1\\ 0 & \text{otherwise,}\end{cases}
\end{equation}
where $a<b$ refers to some fixed enumeration order of $H$ (so that
every unordered conflicting pair is penalized once) and $\lambda>0$ is
a penalty weight, corresponding to the upper-triangular matrix $Q$ of
\eqref{eq:QUBO}.  Note that the isolated contact-graph vertex cut
constraints of the MILP version are not incorporated in the QUBO.

The admissible penalty weights of such QUBO reformulations have been
characterized for the maximum $k$-colorable subgraph problem, which
for $k=1$ is the maximum independent set problem underlying our
models~\cite{Quintero2022}. Our case is in accordance with the more
general results there. Let us discuss the penalty choice in our
present context in detail.

The choice of $\lambda<1$ is inappropriate in general. To show this in
particular, we show an infinite class of Hamming space-minimal
distance pairs, where the corresponding QUBO models will have optimal
solutions that are invalid as a $d$-optimal Hamming-packing.  Consider
binary codes. Note that $N(n;d=d')=2$ iff $\frac{2n}{3}<d'\leq n$; a
proof can be found in \cite{Plotkin_1960,Naszvadi2025}.  If we chose
$n\geq 4$ and $d=(n-1)$, the following set of three codewords
$C=\{\overline{0\ldots 00}, \overline{1\dots 11}, \overline{1\ldots
  10}\}$ realizes Hamming-distances $1$, $d$ and $n$, and only one is
smaller than $d$. Selecting all three words thus contributes $-3$
from the diagonal and a single penalty $\lambda$, giving the objective
value $\lambda-3$. Since every valid packing has objective value at
least $-N(n;d)=-2$, for every $\lambda<1$ this infeasible selection
beats the optimal packing by $1-\lambda>0$:
$\lambda-3=-2-(1-\lambda)<-2$.

For $\lambda>1$, every optimal solution of \eqref{eq:qubomodel1} is a
feasible optimal packing, since a single violated pair can never be
compensated by the linear reward of its two endpoints; for the
boundary case of $\lambda=1$ the optimal value is still attained, but
it the optimal solutions can correspond to infeasible
configurations. Yet such solutions can be transformed in a fast and
easy way to feasible ones.

Recall that our QUBO determines the maximal independent set (MIS)
cardinality of a given finite simple graph.  For $\lambda=1$, the QUBO
in~\eqref{eq:quboQ} in its upper triangular form has a matrix $Q$
whose diagonals are all-$(-1)$-s, and edges are represented as
$(+1)$s; all the other matrix elements are zero.  Denote by $G$ the
conflict graph of the problem; a simple graph whose vertices $V(G)$
are the codewords, and there are edges between the conflicting
pairs. $Q$ is the adjacency matrix of $G$ and we are in search for a
MIS, and the optimal objective value is the MIS cardinality. Let us
show that if the QUBO solution vector corresponds to an invalid
independent set, it can be converted to a valid one in polynomial
time.

Let $S\in \mathbb{R}$ be the optimal solution value, and let the (not
necessarily MILP-feasible) QUBO solution vector be
$\vec{x}=(x_1, x_2,\ldots,x_{|V(G)|})\in \{0,1\}^{|V(G)|}$.  If no two
selected vertices are adjacent, then the solution is feasible.
Otherwise suppose that $(j,k)$ is an edge, and $x_j=x_k=1$.  Now pick
one of them arbitrarily, and change its value from $1$ to $0$.  With
this operation, one selected vertex is removed, and the number of
selected edges has been decremented strictly.  The only remaining
thing to check is the objective value.  By setting it to zero, one
diagonal element is removed after substituting the modified $\vec{x}$
to the QUBO, and at least one off-diagonal element is also removed.
So the objective value $S$ becomes $(S+1-t)$, where $t\geq 1$ is the
number of affected off-diagonals.  Because of the optimality of the
original $\vec{x}$, the objective function cannot be smaller than $S$,
so $t$ must be $1$, so the objective value is unaffected by this step.
Because the number of violating edges is finite, and is upper bounded
by $\binom{|V(G)|}{2}$, iterating the previous process will obtain a
maximum independent set, with cardinality $-S$.  (As a side remark,
this argument applies to any simple graph, not only to the
Hamming-graphs which are under consideration here.)

The just derived quadratic bound can be further refined.  During the
described postprocessing, at the step when $t$ is determined, no
selected edge has either any selected neighbors, nor edges or
vertices, so the number of the postprocessing steps can be upper
bounded with $\lfloor \frac{|V(G)|}{2}\rfloor$, which is a linear
bound.

Finally, let us briefly compare our approach in detail with the one
in~\cite{yukiyoshi2022quantumsearchalgorithmbinary}, the model
conceptually closest to ours. There, binary constant-weight codes are
considered, i.e.\ codes in $H=\mathbb{Z}_2^n$ all of whose words have
the same Hamming weight $w$. As in \eqref{eq:qubomodel1}, a binary
variable is assigned to every admissible word, and the number of
variables is reduced by fixing one codeword and removing all words at
distance less than $d$ from it, which is the analogue of our prune
list $P$. The constant weight allows the Hamming distance to be
expressed through the inner product,
$d_H(a,b)=2(w-\langle a,b\rangle)$, and this is exploited in the
objective function: instead of penalizing conflicting pairs, the
authors minimize $\sum \langle a,b\rangle^{l}x_a x_b$ over all pairs
of words, with an exponent $l$ chosen so that codes with minimum
distance at least $d$ have a lower objective value than any code
violating it. The size of the code is not maximized but prescribed: a
quadratic penalty $\rho(\sum_a x_a-M)^2$ fixes the number of codewords
to a given $M$, so that the model decides whether a code with $M$
codewords exists. Notably, they also consider a given single codeword,
chosen on the basis of symmetry considerations, selected by default.

The resulting QUBO differs from \eqref{eq:qubomodel1} in two respects
relevant for benchmarking. The cardinality penalty couples all pairs of
variables, so that the QUBO matrix is fully dense, whereas the nonzero
off-diagonal elements of our $Q$ correspond to the edges of the
conflict graph. Moreover, the optimal value of \eqref{eq:qubomodel1}
equals $-N(H;d)$ directly, whereas there, a solution certifies the
existence of a code only if its objective value is below a known
threshold.

The QUBO model in~(\ref{eq:qubomodel1},~\ref{eq:quboQ}), together with
\eqref{eq:milpmodel1}, is what the model generator described in
Section~\ref{sec:hammingqubobenchmark} produces for each benchmark
instance.

\section{Hamming-QUBO benchmark}
\label{sec:hammingqubobenchmark}

This Section describes the actual benchmark instances, and positions
them among other QUBO benchmarks.

\subsection{QUBO benchmark generator and benchmark files}

We have developed a model generator script to create our optimal
Hamming packing models. It operates on a Hamming space with the given
parameters, constructs the mutex structure, and builds the desired type
of model: for QUBO output it introduces off-diagonal penalties as in
Eq.~\eqref{eq:qubomodel1}; for linear constrained integer program
generation, it adds linear constraints realizing a MILP model as in
Eq.~\eqref{eq:milpmodel1}. The parameters common to every output type are
the generated model syntax (QUBO, standard MILP etc.), the alphabet cardinalities, and the minimal Hamming distance,
which defaults to $3$.

Beyond these, several optional parameters control the constraint families
of Section~\ref{sec:contactgraphs}: a prune list $P$
(Section~\ref{sec:optmodels}) of ball centers, together with an output
filename that is otherwise guessed from the space parameters; the
no-isolated-vertex cut of Eq.~\eqref{eq:noisolated}; a dual bound, which
for reference MILP models can be set automatically to the
Komamiya--Joshi--Singleton bound; a QUBO penalty weight
(Eq.~\eqref{eq:qubomodel1}); and, for constant-weight word packings, a
restriction along the lines
of~\cite{yukiyoshi2022quantumsearchalgorithmbinary}. A worked example, tracing
one of the smallest instances from the definitions to the literal MILP,
QUBO and label files produced by the generator, is given in Online
Resource~1.

Using the generator script, we have generated $133$ QUBO benchmark
instances, which are included in the research data.  The files are
named with a rational nomenclature specifying the Hamming-space, the
distance $d$, and the removed balls.  For example, for {\tt
  b43322\_4\_0\_qubo.csv} the alphabet cardinalities are
$11,4,3,3,2,2$; the distance is $d=4$, with the $000000$ ball removed.
In what follows we will also refer to these instances with their
basenames, i.e. without the suffix {\tt \_qubo.csv}.
  
We have subdivided the benchmark set to "tiny" and "small"
instances. The "tiny" ones, with up to 300 variables are easy to solve
and were mainly intended for validation. The "small" ones have up to
$1035$ variables and, as it will be clear from the numerical analysis,
contain hard problems, too.  The particular instances are tabulated in
Online Resource~1, and those relevant for the present discussion also in
Tab.~\ref{tab:milp-instances}, with the size and matrix density
(i.e. the percentage of nonzero elements in the main diagonal and the
upper triangle), and the respective optimal values. (The table
contains also MILP timings that will be discussed later in
Section~\ref{sec:lp}.)

\subsection{Comparison with other benchmarks}
\label{sec:instance_features}

In order to investigate the relation of our proposed benchmark set to
other existing benchmarks, we position them in the feature space
defined in the MQLib study~\cite{Dunning2018}. We follow the idea of
instance space analysis~\cite{SmithMiles2023}: every instance is
described by a vector of numerical \emph{features}, and instances with
similar feature vectors are regarded as structurally similar.  As this
feature set was designed to predict which classical Max-Cut/QUBO
heuristic performs best on a given instance, such an analysis reveals
to some extent which are the problems similar to each other from this
perspective.

The features themselves are computed on the weighted graph of the
equivalent Max-Cut problem and summarize its size, density, degrees,
edge weights, clustering, core numbers, and greedy coloring and
independent-set bounds.  QUBO instances are converted to Max-Cut by
the standard reduction, which adds one vertex to the graph of the
QUBO~\cite{Hammer1965}, as in MQLib; Ising models without external
field are Max-Cut problems already and are measured as such, as in the
MQLib library itself.  Of the 58 features
we use 54; three spectral estimates proved numerically unreliable and
the degree assortativity overflows on the largest inputs. The features
are evaluated with the same MQLib version as the one which will be
used in Section~\ref{sec:mqlib}, too.

As the main baseline, we use the MQLib instance library itself: a
collection of over 3000 Max-Cut and QUBO benchmarks. We have also
collected about 5000 instances specifically designed for quantum
devices or quantum-related classical heuristics. They are tabulated in
Table~\ref{tab:benchmark_proximity}, and will be referred to as
``quantum benchmarks'' in what follows.

Features have different units and ranges, so each is first
\emph{standardized}: shifted and scaled to mean zero and unit standard
deviation over the MQLib library. A difference of one unit in a
feature thus corresponds to the typical spread of that feature among
MQLib's graphs. The distance between two instances is the Euclidean
distance of their standardized 54-component feature vectors. For an
instance and a collection, the \emph{nearest-neighbor distance} is
the distance to the most similar member of the collection. MQLib's own
nearest-neighbor distances serve as the yardstick: half of the MQLib
graphs have a neighbor in the library within distance 1.2, and 99\,\%
within 4.5. An instance whose nearest MQLib graph is farther than 4.5
is therefore more remote from the library than 99\,\% of the library's
members are from each other: it lies in a region that MQLib populates
sparsely, if at all. Eight of the features, the statistics of the
local clustering coefficient, are estimated by MQLib from a small
random sample of vertices (21 for a graph of 1000 vertices), drawn by
a fixed but platform-dependent pseudo-random sequence. Measuring our
instances with a different sample moves a typical one by 0.8 in this
space and medians of the distances by up to about 0.2. We therefore quote
distances to one decimal; all three collections are measured with the
same sampling, the one that reproduces MQLib's published feature
values.

\begin{figure}[!htb]
  \centering
  \includegraphics[width=0.92\textwidth]{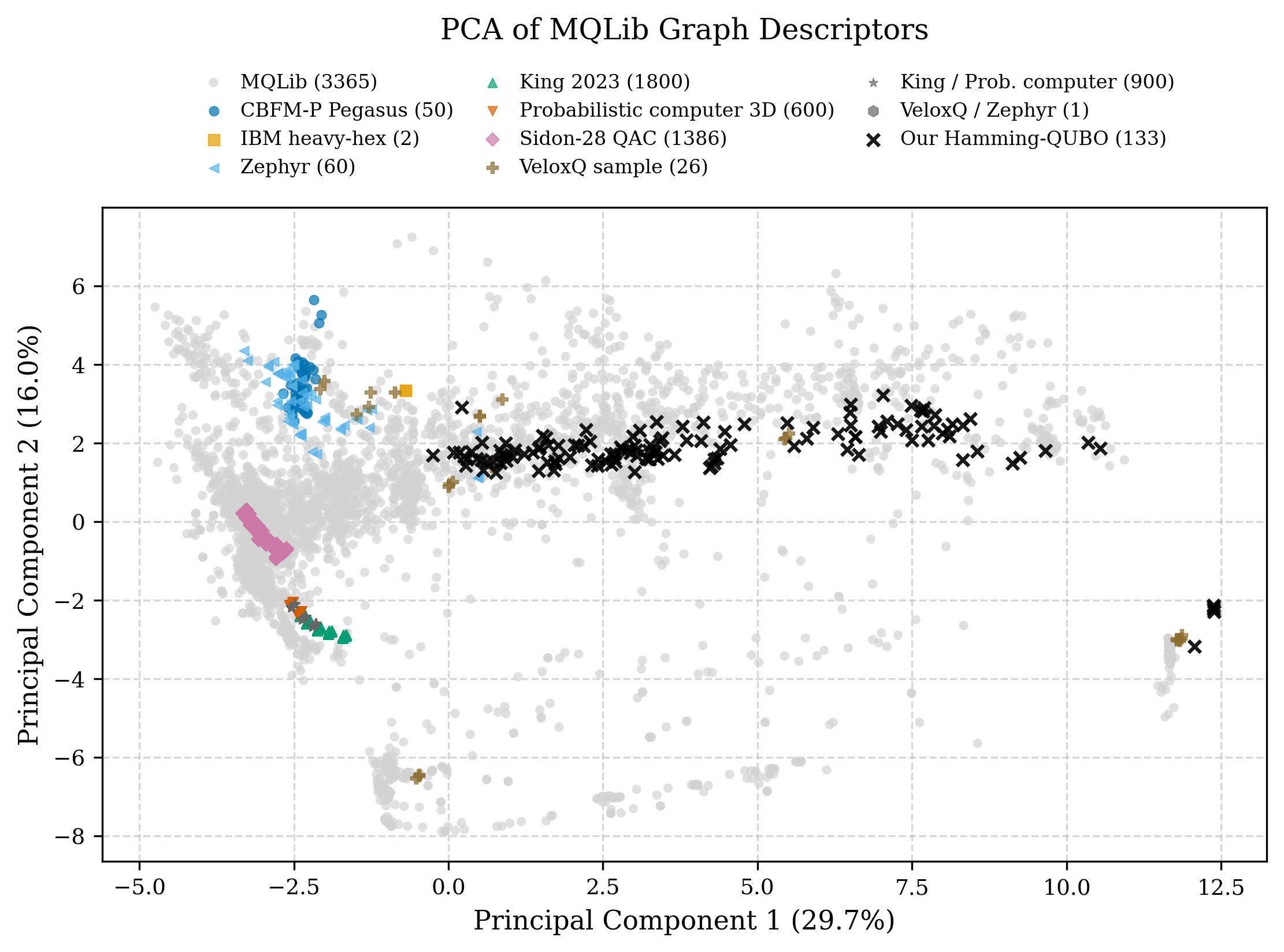}
  \caption{MQLib, quantum benchmarks and Hamming instances projected onto the two principal
  components of the MQLib features. PC1 measures density, PC2 structural regularity
  (negative: regular graphs).}
  \label{fig:cluster_pca}
\end{figure}

To visualize the 54-dimensional space, Fig.~\ref{fig:cluster_pca}
shows the projection onto the two directions along which the MQLib
graphs vary most (principal component analysis, PCA~\cite{Jolliffe2002}, fitted on MQLib
only; the other instances are projected onto the same plane). The
plane retains 46\,\% of the variation of the MQLib features, so it
serves orientation; all distances are computed in the full space. The
two axes have a simple meaning. The first principal component (PC1,
30\,\% of the variation) measures \emph{density}: it weighs the levels
of degree, core number and neighbor degree of the Max-Cut graph
almost equally. The second (PC2, 16\,\%) measures \emph{structural
  regularity} of the Max-Cut graph: it is negative for regular graphs,
in which every vertex has the same degree and core number, and
positive for graphs with skewed, heavy-tailed degree distributions. A
family of graphs with one structural pattern but different densities
therefore forms a horizontal line. MQLib's own regular graphs are an
example: all 334 MQLib graphs below PC2${}=-5$ are regular, and they
spread along PC1 according to their density.

Let us first consider where the quantum benchmarks lie in this space
reflecting the view of classical heuristics.  They are at the edge of
the region populated by MQLib, at its sparse end; see
Table~\ref{tab:benchmark_proximity}.  Their median nearest-neighbor
distance to MQLib, 4.4, is larger than 98.6\,\% of MQLib's internal
nearest-neighbor distances, so a typical quantum benchmark is less
similar to any MQLib graph than almost all MQLib graphs are to each
other. Yet, while 90\,\% of them lie beyond MQLib's 95th percentile
(3.5), only 27\,\% lie beyond the 99th: most quantum benchmarks are
about as far from the library as the few percent most isolated MQLib
graphs are from the rest of it. They border on the library rather than
lie far outside it. Apart from the deliberately varied VeloxQ sample,
all families are sparse and fall into two groups that reflect their
physics. Hardware-native instances with local fields (CBFM-P, Zephyr,
heavy-hex) lie among heterogeneous sparse graphs at positive PC2;
CBFM-P is the most remote family, entirely beyond the 99th
percentile. Spin glasses without field (King et al., the probabilistic
computer, Sidon-28) lie next to MQLib's sparse Steiner-tree and grid
graphs and its two-dimensional toroidal grids; the two-dimensional
tile-planting instances of the VeloxQ sample are as close to MQLib's
two-dimensional toroidal spin glasses as MQLib's graphs typically are
to each other. Each family occupies a small patch, because its
instances share one construction.

\begin{table}[!htb]
  \caption{Nearest-neighbor distances to the MQLib library in the 54-dimensional
  standardized feature space. Median over the instances of each collection; the next
  column gives the fraction of instances farther from MQLib than 99\,\% of MQLib's graphs
  are from their own nearest neighbor (distance 4.5). The first row is MQLib itself
  (each graph to its nearest other graph). Source families overlap (900 instances are
  shared by King et al.\ and the probabilistic computer).}
  \label{tab:benchmark_proximity}
  \centering
  \small
  \begin{tabular}{lrrr>{\raggedright\arraybackslash}p{3.4cm}}
    \toprule
    Collection & $N$ & Median & Beyond & Nearest MQLib graphs \\
    \midrule
    MQLib library~\cite{Dunning2018} & 3365 & 1.2 & 1\,\%   & --- \\
    CBFM-P Pegasus~\cite{Tasseff2024} &   50 & 6.3 & 100\,\% & Beasley~\cite{Beasley1998}, G-set~\cite{Helmberg2000}, GKA~\cite{Glover1998} \\
    IBM heavy-hex~\cite{Chandarana2025} &    2 & 3.7 & 0\,\%   & GKA~\cite{Glover1998} \\
    Zephyr~\cite{Pawlowski2025veloxq,Pawlowski2026} &   61 & 4.5 & 48\,\%  & Beasley~\cite{Beasley1998}, GKA~\cite{Glover1998} \\
    King et al.\ 2023~\cite{King2023} & 2700 & 4.5 & 35\,\%  & SteinLib~\cite{Koch2001} \\
    Probabilistic computer~\cite{Chowdhury2025} & 1500 & 4.5 & 43\,\%  & SteinLib~\cite{Koch2001} \\
    Sidon-28 QAC~\cite{MunozBauza2025} & 1386 & 3.9 & 0\,\%   & 2D toroidal grids~\cite{Dunning2018} \\
    VeloxQ sample~\cite{Pawlowski2025veloxq} &   27 & 4.8 & 52\,\%  & various \\
    All quantum benchmarks   & 4825 & 4.4 & 27\,\%  & \\
    Hamming instances        &  133 & 8.7 & 93\,\%  & GKA~\cite{Glover1998}, Beasley~\cite{Beasley1998} \\
    \bottomrule
  \end{tabular}
\end{table}

Our instances occupy a region that MQLib leaves almost empty. Their
median nearest-neighbor distance to MQLib is 8.7, and 93\,\% of them
lie beyond MQLib's 99th percentile. The emptiness is not only a matter
of the single nearest graph: the average distance to the five nearest
MQLib graphs is larger for a typical Hamming instance than for
99.9\,\% of the MQLib graphs. The closest MQLib graphs are the GKA~\cite{Glover1998} and
Beasley~\cite{Beasley1998} instances, which, like ours, were
originally QUBO problems. In the Max-Cut form of a QUBO, the additional
vertex is joined to every variable whose diagonal entry is not
canceled by the off-diagonal entries of its row; in these instances,
as in ours, this holds for (nearly) all variables, so the graph
contains one vertex adjacent to all others. Such a vertex is rare among
MQLib's other graphs (3\,\%). At the same time,
the Hamming instances are close to each other: the median distance to
the nearest other Hamming instance is 0.8, below MQLib's 1.2, so they
form a coherent family rather than scattered outliers. They are far
from the quantum benchmarks (median 9.2); the reason will become
clear in Section~\ref{sec:qpu}. Only a few small, dense instances of
the VeloxQ and Zephyr samples come that close; all other quantum
families are at distances of about 15--16, at the opposite, sparse end
of the density axis.

In Fig.~\ref{fig:cluster_pca}, 128 of the 133 instances form a
horizontal line that spans a large part of the density range of
MQLib. Along the line, the position is set by the density of the
conflict graph; the generator parameters move an instance along the
line through the density they produce. The height of the line is fixed
by a structural signature that comes from the problem itself, since a
packing objective is not symmetric under complementing all variables:
every diagonal entry is $-1$ and is not canceled by the off-diagonal
entries of its row, so the additional vertex is joined to all other vertices,
and its edges are the most negative ones. Only 1.5\,\% of MQLib's graphs share this combination of
features. The remaining five instances, in the lower right corner,
have complete conflict graphs; together with the additional vertex they
are complete graphs, a region also occupied by complete graphs of
MQLib and by six dense instances of the VeloxQ sample.

\section{Numerical results}
\label{sec:benchmarking}

In this Section we present our computational results using various solvers.

\subsection{Baseline: MILP}
\label{sec:lp}

First we present solutions to the equivalent MILP formulations of each instance, as presented in Eq.~\eqref{eq:milpmodel1}. The MILP files follow the same naming convention as the QUBO instances, and  are available in the research data. These solutions are important because they are exact, thus serving as a baseline for the further analysis and also for the validation of the QUBO formulations. 

The calculations were performed on a single machine equipped with an Intel Core i5-1340P CPU (13th generation, 12 cores -- 4 performance + 8 efficiency -- 16 threads, up to 4.6 GHz) and 62 GB of RAM, running  Devuan GNU/Linux 5 ("daedalus", Linux kernel 6.1). The linear programs were solved using the SCIP Optimization Suite, version 9.2.1, with SoPlex 7.1.3 as the underlying MILP solver~\cite{BolusaniSCIP2024}. The solver is single-threaded. The results for those instances which have properties relevant for the present discussion are tabulated in Tab.~\ref{tab:milp-instances}; the results for all instances are given in Online Resource~1.

{\small 
\begin{longtable}{lrrrrr} 
\caption{Selected instances with MILP benchmark reference features. Only those instances are listed which have properties relevant for the present discussion; the complete list is given in Online Resource~1. The optimal values come from the MILP solutions. The lines typed in boldface are the instances of Table~\ref{tab:top10}, for which the largest number of MQLib algorithms failed. The italic lines are those for which SBM has not found the optimum. Instances marked with an asterisk are those which were addressed with D-Wave quantum hardware.} \label{tab:milp-instances} \\
\toprule 
basename & size & optimum & LP runtime & LP ptime & QUBO density \\
         &      &         & (s)        & (s)      & (\%)         \\
\midrule
\endfirsthead
\toprule 
basename & size & optimum & LP runtime & LP ptime & QUBO density \\
         &      &         & (s)        & (s)      & (\%)         \\
\midrule
\endhead
\midrule 
\multicolumn{6}{r}{\textit{continued on next page}} \\
\midrule 
\endfoot 
\bottomrule 
\endlastfoot 
\multicolumn{6}{l}{\textbf{tiny}} \\
\midrule
\code{{*33333\_3\_0}} & 192 & 18 & 1.43 & 1.13 & 23 \\
\textit{{43332\_3\_0}} & \textit{166} & \textit{14} & \textit{9.91} & \textit{4.07} & \textit{25} \\
\textit{{44332\_3\_0}} & \textit{229} & \textit{18} & \textit{6.04} & \textit{6.03} & \textit{22} \\
\textit{{522222\_3\_0}} & \textit{120} & \textit{11} & \textit{1.05} & \textit{1.05} & \textit{28} \\
\textbf{{533222\_3\_0}} & \textbf{301} & \textbf{21} & \textbf{1202.81} & \textbf{11.32} & \textbf{17} \\
\code{{*722222\_3\_0}} & 172 & 14 & 1.15 & 0.27 & 25 \\
\multicolumn{6}{l}{\textbf{small}} \\
\midrule
\textit{\textbf{{433322\_3\_0}}} & \textit{\textbf{371}} & \textit{\textbf{27}} & \textit{\textbf{2851.54}} & \textit{\textbf{350.40}} & \textit{\textbf{15}} \\
\textbf{{4422222\_3\_0}} & \textbf{451} & \textbf{32} & \textbf{23.99} & \textbf{23.97} & \textbf{12} \\
\textbf{{4432222\_4\_0}} & \textbf{542} & \textbf{13} & \textbf{2540.64} & \textbf{107.71} & \textbf{32} \\
\textbf{{*443322\_3\_0}} & \textbf{505} & \textbf{36} & \textbf{283.11} & \textbf{283.08} & \textbf{13} \\
\textit{{443332\_4\_0}} & \textit{590} & \textit{14} & \textit{1245.96} & \textit{563.76} & \textit{35} \\
\textbf{{444222\_3\_0}} & \textbf{442} & \textbf{32} & \textbf{39.10} & \textbf{39.08} & \textbf{14} \\
\textbf{{543322\_3\_0}} & \textbf{639} & \textbf{36} & \textbf{105.50} & \textbf{105.46} & \textbf{11} \\
\textit{\textbf{{6322222\_3\_0}}} & \textit{\textbf{508}} & \textit{\textbf{32}} & \textit{\textbf{88.28}} & \textit{\textbf{88.25}} & \textit{\textbf{12}} \\
\textbf{{633222\_3\_0}} & \textbf{365} & \textbf{24} & \textbf{4.07} & \textbf{4.06} & \textbf{16} \\
\textbf{{633322\_3\_0}} & \textbf{569} & \textbf{36} & \textbf{54.30} & \textbf{54.26} & \textbf{12} \\
\textit{{6522222\_4\_0}} & \textit{670} & \textit{11} & \textit{429.16} & \textit{80.17} & \textit{33} \\
\textbf{{6665\_3\_0}} & \textbf{925} & \textbf{30} & \textbf{118.58} & \textbf{118.47} & \textbf{15} \\
\textbf{{73333\_3\_0}} & \textbf{480} & \textbf{27} & \textbf{13.49} & \textbf{13.46} & \textbf{16} \\
\textbf{{7666\_3\_0}} & \textbf{1325} & \textbf{36} & \textbf{4206.21} & \textbf{4206.02} & \textbf{12} \\
\textbf{{82222222\_3\_0}} & \textbf{939} & \textbf{64} & \textbf{128.09} & \textbf{128.03} & \textbf{8} \\
\textbf{{8222222\_3\_0}} & \textbf{441} & \textbf{32} & \textbf{4.60} & \textbf{4.58} & \textbf{14} \\
\end{longtable}
} 

Concerning MILP timings, their value is twofold. On one hand, the complete solution time can be compared with that of the exact QUBO solvers', and will be instructive when evaluating the success rate of the heuristics. As it is already apparent from the table, this empirical hardness of the MILP version correlates with the empirical hardness of the QUBO versions.  On the other hand, the primal time, that is, the time of finding the actual optimum without proving its optimality, can be directly compared with (QUBO) heuristics. These aspects will be analyzed in detail along the presentation of the respective QUBO results.

One of the key observations from Table~\ref{tab:milp-instances} is that the ratio of LP runtime to LP ptime can be very large: proving optimality may take orders of magnitude longer than finding the optimum, in accordance with the NP-hard nature of the problem.
Regarding the QUBO density, the densest models are not necessarily the most difficult ones.

\subsection{Exact QUBO solver experiment: Gurobi}

The MILP formulation always yields the exact optimum but runtime
varies enormously: under 0.01\,s on the smallest instances to
$\approx39\,800$\,s (\code{5432222\_4\_0\_qubo.csv}) and
$\approx21\,200$\,s (\code{53333\_3\_0\_qubo.csv}) on the hardest. It
is instructive to compare this behavior to a high-quality traditional
branch-and-bound based solver.  We opt for Gurobi (version 13.0.3,
linux64 build). We emphasize that this comparison is not between
solvers but between the model formulations. We do not benchmark SCIP
or Gurobi against each other or any other exact solver, and the Gurobi
results are not comparable with the heuristics as this is the only
exact QUBO solver used here.

This solver was run on a machine with the same type of CPU as the one
used for the MILP and MQLib benchmarks (Sec.~\ref{sec:lp}), an Intel
Core i5-1340P (16 threads), 62 gigabytes of RAM, under Debian
GNU/Linux 13 (``trixie''). All 16 threads were used, cut generation
was set to aggressive (\code{Cuts=3}), and all other parameters were
left at their defaults. As the two solvers address different models,
the comparison below is one of the QUBO and MILP formulations rather
than of the solvers. Gurobi had more parallel resources than the
single-threaded MILP solver, so this does not weaken the conclusions
drawn from it. We also note that the presolver of Gurobi linearized
21 of the instances.

The solver was run with a 3600\,s time limit, which is higher than the
time the MILP solver needed, with the exception of the few instances
that were the hardest for the latter.  It proved optimality only on 40
of the 133 instances (30\,\%); the remaining 93 hit the time limit,
although its best solution coincided with the MILP optimum on all 133
instances. The optimum was in fact found within 3\,s on every instance,
on 117 of them already by the heuristics run before presolve; the
difficulty lies mainly in the bound. On the instances that hit the
time limit, the final relative gap between the incumbent and the bound
ranged from 31\,\% to 6853\,\%, with a median of 814\,\%.
From this one can conclude that these QUBOs are indeed a challenge for
an exact solver, and, as expected, using the MILP version of the models
with the appropriate solver is more efficient in the case of exact
solvers. However, it will be apparent that QUBO heuristics are
competitive in efficiency.

\subsection{Traditional heuristics: MQLib}
\label{sec:mqlib}

The MQLib experiment was performed on the same computer as the MILP
calculations, with MQLib built from source from the \code{MQLib/MQLib}
GitHub repository (no versioned releases exist; commit \code{5854962},
2025-11-12). It implements 12 Max-Cut and 27 QUBO heuristics. Each (instance,
heuristic) pair was run once with a 20\,s wall-clock time budget. The time budget was chosen based on experimentation; it is far below the wall-clock time the exact solvers needed.

\begin{figure}[p]
  \centering
  \includegraphics[width=\textwidth]{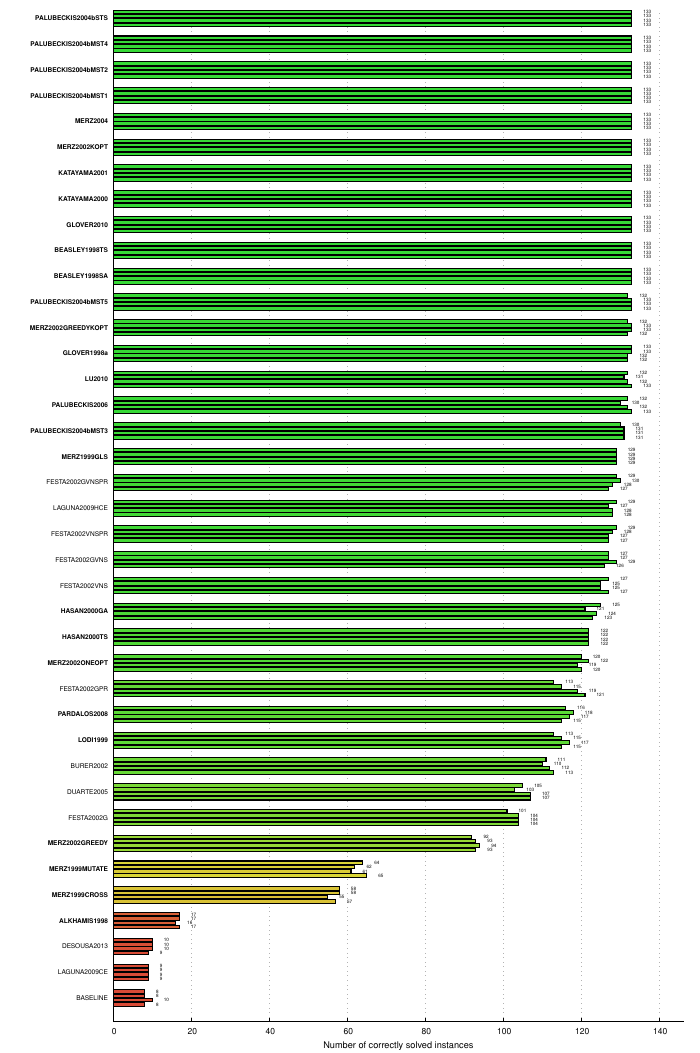}
  \caption{Per-algorithm optimality rates across the 39 MQLib
    algorithms on 133 instances, sorted by rate. The 27 QUBO
    algorithms' name is typed in bold. The four bars for each
    algorithm are for different seeds, from top to bottom: 42,
    137, 8312, and 38013.}
  \label{fig:gap_dist}
\end{figure}

For each of the 133 instances, there is a number of algorithms
capable of finding the optimal solution within the given time budget.
Figure~\ref{fig:gap_dist} summarizes the algorithms' performance.
We did not aim at a detailed study of seed dependence, yet we have
performed the experiments with 4 different ad-hoc seeds. This suggests
that the seed dependence is moderate, and the results for all the four
seeds are similar. We give details for seed value of $42$ in what follows.

It appears that QUBO heuristics typically perform better than Max-Cut heuristics. While the best-of-39 ensemble finds the optimum on every instance, the
three weakest algorithms succeed on only 6--46\,\% of instances.
Instance difficulty tends to grow with size, though not strictly: \code{82222222\_3\_0\_qubo.csv} ($|H'|=939$, optimum~$-63$), one of the three instances that defeat the largest number of algorithms (22 of the 39, 12 of which are QUBO heuristics), is among the larger ones, whereas \code{433322\_3\_0} ($|H'|=371$) is equally hard.

\begin{figure}[h]
  \centering
  \includegraphics{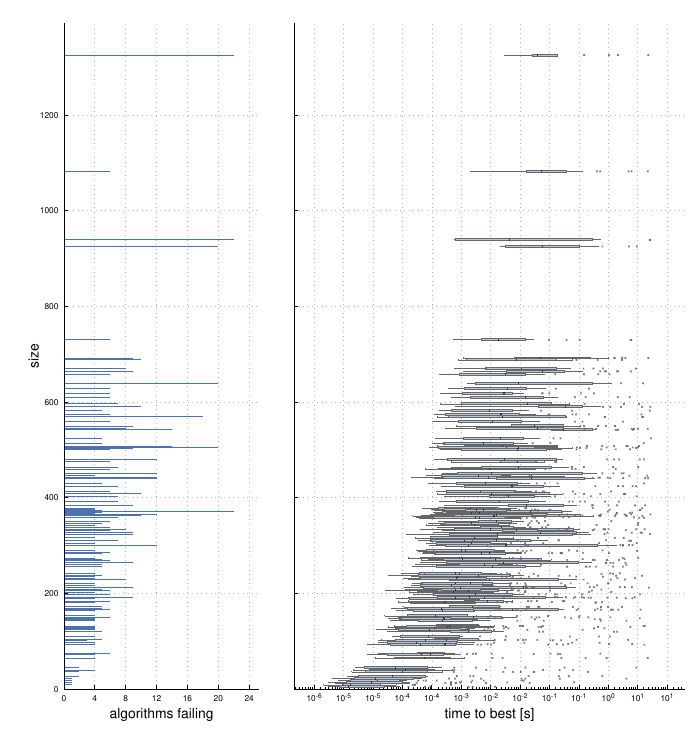}
  \caption{MQLib heuristic performance (seed 42) on the 133 benchmark instances, as a
function of instance size (vertical axis, shared by both panels, with
instances arranged bottom to top by increasing size). Left: number of the
39 MQLib heuristics that failed to reach the instance's known optimum
(obtained from the SCIP-MILP formulation), drawn as a horizontal stem from
zero to the failing count. Right: distribution of the wall-clock time to
the best solution, taken over only the heuristics that did reach the
optimum for that instance; each box spans the first to third quartile
with a vertical line at the median, whiskers extend to the most extreme
value within 1.5 times the interquartile range, and values beyond the
whiskers are shown individually (x markers); note the logarithmic time
axis. A few instances share the exact same size; these are offset
slightly on the (shared) size axis so their stems/boxes remain
distinguishable.
}
  \label{fig:mqlibtiming}
\end{figure}

Figure~\ref{fig:mqlibtiming} shows per-instance algorithm failure
counts and the time-to-best for algorithms reaching the optimum as a
function of instance size $n$. Failure counts tend to grow with $n$,
but there are smaller instances, too, on which more algorithms fail. As for the time-to-best behavior, the bigger
problems are apparently more challenging, albeit the per-instance
median remains near zero, meaning that the algorithms that do succeed
find the optimum quickly, regardless of instance size. From the
presence of outliers it can be seen that even the small instances
tend to be challenging for some of the algorithms.
The instances that presented the biggest challenge for MQLib, namely the top ten together with those tied with the tenth, are tabulated in Tab.~\ref{tab:top10}. (These are the ones set in boldface
in Tab.~\ref{tab:milp-instances}.)
\begin{table}[h]
  \centering
  \caption{Instances with the highest number of MQLib algorithms failing to
    reach the MILP optimum with seed=42.}
  \label{tab:top10}
  \small
  \begin{tabular}{lrr}
    \toprule
    Instance & $n$ & Failing \\
    \midrule
    \code{82222222\_3\_0} & 939 & 22 \\
    \code{433322\_3\_0} & 371 & 22 \\
    \code{7666\_3\_0} & 1325 & 22 \\
    \code{443322\_3\_0} & 505 & 20 \\
    \code{543322\_3\_0} & 639 & 20 \\
    \code{6665\_3\_0} & 925 & 20 \\
    \code{633322\_3\_0} & 569 & 18 \\
    \code{4432222\_4\_0} & 542 & 14 \\
    \code{6322222\_3\_0} & 508 & 14 \\
    \code{4422222\_3\_0} & 451 & 12 \\
    \code{444222\_3\_0} & 442 & 12 \\
    \code{533222\_3\_0} & 301 & 12 \\
    \code{73333\_3\_0} & 480 & 12 \\
    \code{633222\_3\_0} & 365 & 12 \\
    \code{8222222\_3\_0} & 441 & 12 \\
    \bottomrule
  \end{tabular}
\end{table}

One can find that for each instance there are traditional algorithms
that find the optimum efficiently.  In fact in the case of certain
problems they do so even faster than the MILP solver finds the optimum
first (without proving its optimality yet); compare the 20-second time
budget MQLib had with the ``LP ptime'' column of
Tab.~\ref{tab:milp-instances}. This suggests that in the case of this
set of problems, using QUBO heuristics to find an initial solution,
and using it as a starting configuration for the MILP solver with the MILP
version of the problem, thereby orchestrating QUBO heuristics and MILP
solvers would be a good strategy for solving these problems. Apart
from that, the results also illustrate that our benchmark set contains
instances that are not trivial for certain traditional heuristics.

\subsection{Simulated bifurcation machine (SBM)}

We used the discrete version of simulated bifurcation~\cite{SBM} as
implemented in the VeloxQ platform~\cite{Pawlowski2025veloxq}, with
its default settings apart from the following: $4\,096$ samples
(independent trajectories) per instance, and $5\,000$ time steps of
length $\Delta t=0.05$. The same settings were used for all 133
instances, without any per-instance tuning. With these, an SBM run
takes 0.17--2.25\,s per instance.  This wall time, however, is not related to
the hardness of the given instance: the time to simulate the given
number of time steps for a sample and the number of samples determines
this time regardless of the problem structure, the small variance is
due to technical artifacts.

SBM finds the exact optimum in 126 of 133 instances (95\,\%).  The
seven sub-optimal cases are listed in Table~\ref{tab:sbm_miss}. Given
the short wall time required for this solution one can conclude that
SBM performs very well on these instances, and even provides a fair
suboptimal solution for the handful failing ones.
\begin{table}[h]
  \centering
  \caption{Instances where SBM does not reach the MILP optimum.}
  \label{tab:sbm_miss}
  \small
  \begin{tabular}{lrrr}
    \toprule
    Instance & Optimum & SBM best & Gap \\
    \midrule
    \code{6322222\_3\_0} & $-31$ & $-29$ & 2 \\
    \code{43332\_3\_0} & $-13$ & $-11$ & 2 \\
    \code{44332\_3\_0} & $-17$ & $-15$ & 2 \\
    \code{433322\_3\_0} & $-26$ & $-25$ & 1 \\
    \code{443332\_4\_0} & $-13$ & $-12$ & 1 \\
    \code{6522222\_4\_0} & $-10$ & $-9$ & 1 \\
    \code{522222\_3\_0} & $-10$ & $-9$ & 1 \\
    \bottomrule
  \end{tabular}
\end{table}
Because the underlying MILP is monotone -- removing any codeword from
a feasible packing yields another feasible packing~\cite{Naszvadi2025}
-- the objective in \eqref{eq:qubomodel1} attains every integer
between $0$ and $-N(H;d)$: dropping one selected codeword at a time
from an optimal solution decreases $|C|$, and hence increases the
objective, by exactly $1$ at each step. The small integer gaps
reported for the simulated bifurcation machine
(Table~\ref{tab:sbm_miss}) are therefore genuine near-optimal packings
missing one or two codewords, not artifacts of the penalty encoding.

There seems to be no correlation between the hardest MQLib and hardest SBM instances. 
Because of the relatively small amount of instances that have proven to be hard for SBM, this is not a statistically well-established statement. 
It is interesting though that the number of failing MQLib heuristics varies for the $7$ SBM-hard instances; the number of failing algorithms is 5, 6, 8, 8, 10, 14 and 22 in particular.

\subsection{D-Wave QPU results}
\label{sec:qpu}

In order to understand the behavior of our instances on a quantum
processing unit (QPU), we examine their Ising-equivalent. The
conversion from QUBO form~\eqref{eq:QUBO} to the Ising
form~\eqref{eq:Ising} is done with the substitution
\begin{equation}
x_a=\frac{1+s_a}{2},\quad s_a\in\{-1,+1\}\label{eq:qubo2Ising}
\end{equation}
The couplings (in a symmetric matrix), the local fields, and the offset read
\begin{align}
  J &= \frac18\left(Q+Q^\top\right)-\frac14 \diag(\diag(Q)), \nonumber\\
  \mathbf{h} &= \frac14\left(Q+Q^\top\right)\mathbf{1}, \label{eq:qubo2Isingcoeffs}\\
  c &= \frac14\left(\mathbf{1}^\top Q\,\mathbf{1}+\operatorname{tr}Q\right), \nonumber
\end{align}
where $\mathbf{1}$ is the all-ones vector.
For our QUBOs with elements~\eqref{eq:quboQ}, the Ising version reads
\begin{align}
  J &= \frac{\lambda}{8}\,A, \label{eq:hammingIsingJ}\\
  \mathbf{h} &= -\frac12\,\mathbf{1}+\frac{\lambda}{4}\,A\mathbf{1}, \label{eq:hammingIsingh}\\
  c &= -\frac{|H'|}{2}+\frac{\lambda}{8}\,\mathbf{1}^\top A\,\mathbf{1}, \label{eq:hammingIsingc}
\end{align}
where $A$ is the adjacency matrix of the conflict graph, that is,
$A_{ab}=1$ iff $1\le d_H(a,b)\le d-1$ and $A_{ab}=0$ otherwise. Hence,
$(A\mathbf{1})_a$ is the degree of $a$, and thus, by~\eqref{eq:hammingIsingh},
the degree of each node contributes to the respective local field, with
a coefficient $\lambda/4$.  Meanwhile the off-diagonal elements
in~\eqref{eq:hammingIsingJ} are constant $\lambda/8$, leading to a
coupling of $\lambda/4$.  These have relevant consequences for an analog
device, because of the limited accuracy and numerical ranges.

On one hand, the expression for the local fields shows why it is
important to keep $\lambda$ as low as possible: the $-1/2$, i.e.\ the
first summand of the local field~\eqref{eq:hammingIsingh},
decides whether a node is selected,
and this can become a negligible contribution to the field strength
with growing $\lambda$. This justifies our choice of $\lambda=1$, the
smallest value for which the feasibility can be maintained. The
importance of tight penalty weights for quantum hardware has also been
pointed out for QUBO reformulations of the maximum $k$-colorable
subgraph~\cite{Quintero2022} and the Max-$k$-Cut~\cite{Harkness2025}
problems.

On the other hand, the ratio of the Ising parameters, including both
couplings and local fields, plays an important role.  With $\lambda=1$,
every conflicting pair is coupled with the same strength
$1/4$. Meanwhile, for the local fields~\eqref{eq:hammingIsingh} we get
$h_a=-\frac12+\frac14\deg(a)$, with $\deg(a)=(A\mathbf{1})_a$: the
fields grow with the degree of the conflict graph.  Denoting the
maximum degree in this graph by $\deg_{\max}=\max_a\deg(a)$, the
largest coefficient is thus a local field whenever $\deg_{\max}>4$,
and the ratio of the largest field to the coupling is
\begin{equation}\label{eq:fieldcouplingratio}
  R=\frac{\max_a h_a}{1/4}=\deg_{\max}-2.
\end{equation}
The devices have separate programmable ranges for the fields and the
couplings, $[-h_{\max},h_{\max}]$ and $[-J_{\max},J_{\max}]$, with
$h_{\max}=4$ on the Pegasus and $h_{\max}=6$ on the Zephyr systems used
below, and $J_{\max}=1$ on both. When the largest field is scaled to
$h_{\max}$, a coupling occupies only a fraction $(h_{\max}/J_{\max})/R$
of the coupling range, and its relative accuracy decreases accordingly.

Large fields do not mean, however, that the fields dominate the energy
landscape. Since every word $a$ has $\deg(a)$ couplings of strength
$\lambda/4$, \eqref{eq:hammingIsingh} states that, apart from the
$-1/2$, each field equals the sum of the absolute values of the
couplings acting on its spin. The fields would dominate, and the ground
state would be obtained by aligning every spin against its field, if
each $|h_a|$ exceeded this sum; instead, it falls short of it by
exactly $1/2$, for every variable of every instance. Aligning all spins
against their (positive) fields gives $\mathbf{s}=-\mathbf{1}$, the
empty packing, of energy $0$. What decides the state of a spin is the
effective field acting on it, i.e.\ the coefficient of $s_a$ in the
energy at fixed values of the other spins,
\begin{equation}\label{eq:effectivefield}
  f_a = h_a+\frac{\lambda}{4}\sum_b A_{ab}s_b = -\frac12+\frac{\lambda}{2}\,m_a,
\end{equation}
where $m_a$ is the number of selected neighbors of $a$. It does not
depend on the degree: a word without selected neighbors is favored by
$-1/2$; for $\lambda=1$, a word with one selected neighbor is
indifferent, and one with two or more is driven out of the packing.
The fields, of order $\deg_{\max}/4$, and the couplings thus cancel
almost exactly, and the information that determines the packing is
contained in their small remainder.

The maximum degree, and hence the ratio~\eqref{eq:fieldcouplingratio},
is determined by the parameters of the Hamming space. Without pruning, the conflict graph is regular, every word
having $|B(a,d-1)|-1$ neighbors, where
\begin{equation}\label{eq:ballvolume}
  |B(a,r)|=\sum_{j=0}^{r}e_j(k_1-1,\dots,k_n-1)
\end{equation}
with $e_j$ the $j$-th elementary symmetric polynomial. Pruning can only
remove neighbors, so $\deg_{\max}\le|B(a,d-1)|-1$. For $P=\{\bar0\}$
and $n\ge 2d-1$ equality holds, as every word of weight at least
$2d-1$ keeps its whole ball in the pruned space. This is the case for
$72$ of our $133$ instances. For fixed $d$, $\deg_{\max}$ grows polynomially,
with degree $d-1$, in the length and the alphabet sizes; for a uniform
alphabet of size $k$ it is approximately $\binom{n}{d-1}(k-1)^{d-1}$.
The ratio~\eqref{eq:fieldcouplingratio} therefore grows without bound
with the instance size, reaching $479$ in our set. This is in contrast
with hardware-native benchmarks, whose degree is bounded by that of
the hardware graph.

The high degree is also an obstacle to embedding the problem into the
hardware graph. The qubits of current D-Wave processors have degree
$D=15$ (Pegasus~\cite{Boothby2020}) or $D=20$
(Zephyr~\cite{Boothby2021zephyr}), so a variable with $\deg(a)>D$ has
to be represented by a chain of physical qubits, coupled
ferromagnetically so that they act as a single spin~\cite{Choi2008}.
A chain of $L_a$ qubits is connected, hence it uses at least $L_a-1$
couplers internally, leaving at most $L_aD-2(L_a-1)$ for the conflicts
of $a$, thus
\begin{equation}\label{eq:chainlength}
  L_a\ge\frac{\deg(a)-2}{D-2}.
\end{equation}
Heuristic embeddings~\cite{Cai2014} typically use considerably longer
chains. The local field $h_a$ of~\eqref{eq:hammingIsingh} is distributed
among the $L_a$ qubits of the chain, which relieves the field range,
while the chain coupling has to be strong enough to keep the chain
aligned against the couplings acting on it~\cite{Choi2008}. The chain
coupling then becomes the largest coefficient relative to its range,
and sets the scale of the whole problem. The selection term $-1/2$
of~\eqref{eq:effectivefield} is distributed among the qubits of the
chain as well, so that each of them receives only $-1/(2L_a)$ of it;
moreover, long chains are prone to breaking.

Our instances thus do not fit naturally to D-Wave-style hardware:
their conflict graphs are dense, the local fields are large compared
to the couplings~\eqref{eq:fieldcouplingratio}, and the effective
field~\eqref{eq:effectivefield} that decides the selection of a word is
a small remainder of large terms that cancel.  As they are, however, small and well-structured
problems, they can serve as a stress test for embedding and for the
accuracy of analog devices.

Given these facts, only three of the instances were submitted to the
QPU: \code{722222\_3\_0\_qubo.csv} of size 172 and
\code{33333\_3\_0\_qubo.csv} of size 192 from the ``tiny'' subset, and
\code{443322\_3\_0\_qubo.csv} of size 505 from the ``small''.
The five D-Wave systems that have been used are tabulated in
Table~\ref{tab:qpu_emb}. Each run used \code{num\_reads=500} and
three annealing times: 20, 200, and 1000\,$\mu$s.

In the Ising form~\eqref{eq:hammingIsingJ}--\eqref{eq:hammingIsingc}
with $\lambda=1$, all three instances have a single coupling strength
$1/4$, on $3\,681$,
$4\,080$ and $16\,140$ conflicting pairs, respectively, that is, on
$25\,\%$, $22\,\%$ and $13\,\%$ of all variable pairs. Due to the
pruning, the conflict graphs are not regular: the degrees range over
$35$--$51$, $38$--$50$ and $54$--$70$, so the local fields lie in the
intervals $[8.25,12.25]$, $[9,12]$ and $[13,17]$. All fields are
positive and of the same order of magnitude, and the selection term,
the first summand of~\eqref{eq:hammingIsingh}, amounts to only about
$3$--$6\,\%$ of them. The
ratio~\eqref{eq:fieldcouplingratio} of the largest field to the
coupling is $R=49$, $48$ and $68$. Within our benchmark set this is
on the low side: $R$ ranges from $2$ to $479$ with a median of $151$.
Even for these instances, though, if the largest field were scaled to
$h_{\max}$, a coupling would occupy only $8.2$, $8.3$ and $5.9\,\%$ of
the coupling range on Pegasus, and $12.2$, $12.5$ and $8.8\,\%$ on
Zephyr.

\begin{table}[h]
  \centering
  \caption{D-Wave systems used (Adv~4.1 stands for
    \code{Advantage\_system4.1}, Adv2~1.11 for
    \code{Advantage2\_system1.11}, etc.), with the number of active
    qubits at the time of execution, and the embedding results:
    wall-clock time of the embedding search, physical qubits used and
    overhead factor, and range of chain lengths. A dash means that no
    embedding was found.}
  \label{tab:qpu_emb}
  \small
  \begin{tabular}{lrrrrr}
    \toprule
     & Adv 4.1 & Adv 6.4 & Adv2 1.11 & Adv2 2.1 & Adv2 4.3 \\
    \midrule
    Region        & na-west-1 & na-west-1 & na-west-1 & eu-central-1 & na-east-1 \\
    Architecture  & Pegasus & Pegasus & Zephyr & Zephyr & Zephyr \\
    Active qubits & 5\,627 & 5\,612 & 4\,582 & 4\,516 & 1\,203 \\
    \midrule
    \multicolumn{6}{l}{\code{722222\_3\_0}, $n=172$} \\
    \quad time (s)        & 45 & 82 & 56 & 73 & 150 \\
    \quad physical qubits & 3\,359 & 3\,205 & 2\,909 & 2\,879 & -- \\
    \quad overhead        & $19.5\times$ & $18.6\times$ & $16.9\times$ & $16.7\times$ & -- \\
    \quad chain length    & 8--31 & 8--32 & 6--32 & 7--29 & -- \\
    \midrule
    \multicolumn{6}{l}{\code{33333\_3\_0}, $n=192$} \\
    \quad time (s)        & 156 & 293 & 431 & 379 & 62 \\
    \quad physical qubits & 4\,312 & 4\,413 & -- & -- & -- \\
    \quad overhead        & $22.5\times$ & $23.0\times$ & -- & -- & -- \\
    \quad chain length    & 10--38 & 10--43 & -- & -- & -- \\
    \midrule
    \multicolumn{6}{l}{\code{443322\_3\_0}, $n=505$} \\
    \quad time (s)        & 1\,012 & 1\,008 & 1\,006 & 1\,008 & 516 \\
    \quad physical qubits & -- & -- & -- & -- & -- \\
    \bottomrule
  \end{tabular}
\end{table}

Table~\ref{tab:qpu_emb} describes the quantum annealer devices that
were used for our evaluations, and the results of embedding. The
characteristics of embedding are in line with the expectations
above. Summing the bound~\eqref{eq:chainlength} over the variables,
\code{722222\_3\_0} needs at least $628$ physical qubits on Pegasus
and $456$ on Zephyr, \code{33333\_3\_0} at least $688$ and $496$, and
\code{443322\_3\_0} at least $2\,657$ and $1\,942$; the shortest
admissible chains for the largest-degree variables have $4$ (Pegasus)
and $3$ (Zephyr) qubits for the two tiny instances, and $6$ and $4$
for \code{443322\_3\_0}. The heuristic embedder~\cite{Cai2014} uses
$5$--$6.5$ times as many qubits, with chains of up to $29$--$43$
qubits. No embedding was found for \code{443322\_3\_0} on any of the
systems, so no QPU results exist for this instance. On the smallest
system, with $1\,203$ active qubits, fewer than any of the successful
embeddings used, no instance could be embedded. Zephyr gave more
compact embeddings for \code{722222\_3\_0}, but no embedding of
\code{33333\_3\_0} was found on it.

To quantify the effect of the chains, we have reconstructed the
problems actually programmed on the devices from the embeddings,
following the procedure of the Ocean software used for the runs: fields
and couplings are split evenly over the chains and over the couplers
between two chains, respectively, and all coefficients are scaled
automatically to the programmable ranges. The chain strength, left at
its default (uniform torque compensation), is $2.3$ in the units
of~\eqref{eq:hammingIsingJ}, about nine times the logical coupling. On
all six embeddings, the fields split over the chains stay well inside
the field range (the largest one per qubit is at $10$--$13\,\%$ of
$h_{\max}$), so the scale is set by the chain coupling. A logical
coupling then ends at $10.8\,\%$ of the coupling range, comparable to
the values without chains above, and it is further split over up to
$20$--$26$ physical couplers, with a median per-coupler value of
$5.4$--$10.8\,\%$. The selection term $-1/2$
of~\eqref{eq:effectivefield} is affected the most: divided over the
chain and scaled, it amounts to $0.1$--$0.7\,\%$ of the field range per
qubit. A stronger chain coupling, which could reduce chain breaks,
would compress all logical terms further, since it sets the scale. Poor
QPU output is therefore to be expected. (The scripts computing these
figures from our QPU data are part of the research data.)

\begin{figure}[h]
  \centering
  \includegraphics[width=\textwidth]{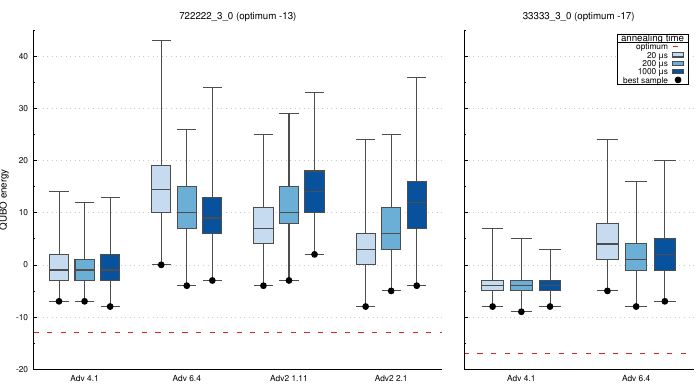}
  \caption{Raw D-Wave QPU sample energies for \code{722222\_3\_0} (left)
    and \code{33333\_3\_0} (right), $500$ reads per system and annealing
    time. Boxes span the first to third quartile, with a tick at the
    median; whiskers extend to the minimum and maximum, and the dot marks
    the best sample. The dashed line is the optimum of the QUBO, which is
    one less than the MILP optimum of Table~\ref{tab:milp-instances}, as
    the fixed word $\bar 0$ is not a variable of the QUBO.}
  \label{fig:qpu}
\end{figure}
The QPU output is far from the optimum on both instances
(Fig.~\ref{fig:qpu}): the best samples have energies $-8$ and $-9$,
against the optima $-13$ and $-17$, respectively. Moreover, the median
sample energy is positive for \code{722222\_3\_0} on three of the four
systems and for \code{33333\_3\_0} on Adv~6.4. Since with $\lambda=1$
the energy is the number of conflicting pairs minus the number of
selected words, the typical sample is infeasible: only $0$--$4.6\,\%$
of the samples of a run are feasible, and the median sample contains
$4$ to $41$ conflicting pairs, depending on the run. Only Adv~4.1
keeps the bulk of the samples around zero or below. A longer annealing
time does not help; on both Zephyr systems the results for
\code{722222\_3\_0} get worse with it. This is in line with the
expectations above.

The coefficient analysis above predicts poor performance from the
structure of the problem alone, and the experiments are in line with
it. They do not, however, isolate this mechanism from others acting in
the same direction. Chains break frequently: every sample contains
broken chains, on average $8$--$17\,\%$ of the chains, depending on
the run, and these are resolved by majority vote, which can create or
remove conflicts. Further possible contributions are the noise and
calibration errors of the analog device, deviations from adiabatic
evolution, and, at $\lambda=1$, the ties
of~\eqref{eq:effectivefield}, which make configurations with isolated
conflicting pairs degenerate with feasible ones. Disentangling these
contributions would require dedicated experiments, e.g.\ varying the
chain strength and $\lambda$, which are beyond the scope of the present
paper. Whichever of them dominates, the instances in their present form
seem less suited for benchmarking the optimization performance of
current quantum annealers than for testing embedding and the accuracy
of analog devices.

\section{Conclusions and outlook}
\label{sec:conclusions}
\label{sect:conclusions}

Starting from the MILP formulation of an important problem in coding
theory, we have proposed a QUBO benchmark problem set. Our aim was to
find small yet nontrivial problems coming from a real application. We
have implemented a generator script and have studied 133 instances in
detail.

In order to compare our problems with other benchmarks, we have
carried out an analysis in the feature space defined in the MQLib
study, thereby reflecting the structural properties of the problems
from the perspective of classical heuristics. A comparison against the
MQLib collection revealed that our Hamming instances form a coherent
family in a region that the collection leaves almost empty, spanning a
wide range of densities and extending well beyond the sparse end where
the quantum benchmarks lie. We have also analyzed a large set of
benchmark problems designed for quantum devices and quantum-inspired
algorithms. Our problems turned out to be different from these. As a
side result, we have found that the quantum benchmarks sample a
narrow, sparse, lattice- and hardware-like border of the region
covered by MQLib.

We have solved the MILP version of the models to obtain a baseline,
and made an experiment with a state-of-the-art QUBO exact solver,
Gurobi. This has been done to compare the two formulations (not the
two solvers). The results suggest that the QUBO formulation is hard
for exact solvers, from the point of view of proving optimality. Our
benchmark instances are therefore good candidates for testing and
actual benchmarking of exact QUBO solvers, which is a possible
direction for future research.

We have studied in detail the behavior of classical heuristics on our
benchmark set. Having the MILP solutions as a stable baseline, we have
found that many, but not all, of the algorithms in the MQLib
collection perform well on our instances, and that the level of
difficulty observed in the MILP version correlates with the level of
difficulty for classical heuristics. We have also tested the instances
with a simulated bifurcation machine, a highly ``quantum-like''
classical solver, and found that it performs uniformly well on these
instances, although our results do not reveal any specifically
quantum-like feature of its behavior. Notably, in several cases the
QUBO heuristics found the optimum in less time than what the MILP
solver needed to even first encounter it. This suggests that a
combined approach, using the solutions of QUBO heuristics as a warm
start for the MILP solver, could be efficient for these problems.

We have analyzed the problems from the perspective of quantum
annealing hardware. Rewriting the instances in Ising form shows that
they are antiferromagnets with large local fields that are almost
exactly balanced by the couplings: the effective field acting on each
spin is of order one, independently of the degree. This
is a property hardly visible from the perspective of classical
algorithms, for which our problems are special maximum independent set
problems. In the Ising form, the ratio of the largest local field to
the coupling grows with the instance size, and the key difficulty is
that the information is encoded in small parameter differences, which
are easily lost when programming a limited-accuracy device; on the
devices we used, the term deciding the selection of a word ends up
below $1\,\%$ of the field range per qubit. On the one
hand, this illustrates the usefulness of viewing a QUBO problem from
the complementary perspectives of classical operations research and
physics. On the other hand, together with the strong connectivity of
the Ising graphs, it explains why they differ so much from known
quantum benchmarks: in spite of their small size, they are not
suitable for the kind of quantum devices we have tested them on, in
line with the experiments described in this paper, although chain
breaks and other device effects, not disentangled here, may also
contribute. Other
architectures, such as Rydberg atom arrays~\cite{Ebadi2022}, may prove
more suitable. It would also be instructive to examine these problems
on gate-based machines with the QAOA~\cite{Farhi2014,Blekos2024} or
VQE~\cite{Peruzzo2014} algorithms, or with gate-annealing
combinations~\cite{Deshpande2026}.

We also remark that decomposing larger instances by pruning further
codewords yields subproblems of a size that can be embedded on current
annealers. We conducted preliminary experiments not detailed here, in
which 82 such subproblems with 104--190 variables could be embedded
and sampled. The pruning further decreases the regularity of the
conflict graph. However, it leaves the maximum degree and the density,
and hence the ratio of the local fields to the couplings, essentially
unchanged, and the sample quality remained comparable to that of the
instances studied here. Whether symmetry breaking itself improves QPU
suitability remains an open question.

\begin{acknowledgements}
  The authors gratefully acknowledge the J\"ulich Supercomputing
  Centre (\url{https://www.fz-juelich.de/jsc}) for funding this project
  by providing computing time on the D-Wave Advantage\texttrademark{}
  System JUPSI through the J\"ulich UNified Infrastructure for Quantum
  computing (JUNIQ).
\end{acknowledgements}

\section*{Declarations}

\paragraph{Funding.}

This research was co-funded by the Slovenian Research and Innovation Agency ARIS through the research project Quantum Solver for Hard Binary Quadratic Problems (Grant No. J7-50186) and through the annual work programme of Rudolfovo; and by the
QEC4QEA project (Grant Agreement No. 101194322), funded by the European High Performance Computing Joint Undertaking (EuroHPC JU) and participating countries, including Slovenia, under the Horizon Europe research and innovation programme.

The project was supported by Ministry of Culture and Innovation and
the National Research, Development and Innovation Office within the
Quantum Information National Laboratory of Hungary (Grant
No. 2022-2.1.1-NL-2022-00004).

\paragraph{Competing interests.}
The authors have no relevant financial or non-financial interests to
disclose.

\paragraph{Data availability.}
The benchmark instances (QUBO and MILP files), the generator script,
and the raw results underlying this study are openly available
at~\url{https://gitlab.wigner.hu/naszvadi.peter/hammingqubobenchmarks}

\paragraph{Code availability.}
The model generator and the analysis scripts are available in the same
repository as the data.

\paragraph{Author contributions.}
P\'eter Naszvadi: Conceptualization; Methodology; Software; Investigation; Formal analysis; Writing -- review \& editing.
Marcell Tukora: Software; Investigation; Formal analysis (comparison with other benchmarks)
Roman Ku\v{z}el: Software; Investigation (quantum annealing)
M\'aty\'as Koniorczyk: Conceptualization; Methodology; Software; Investigation; Formal analysis; Writing -- original draft; Writing -- review \& editing
\paragraph{Use of generative AI.}
During the preparation of this work the authors used Claude
(Anthropic) for language editing, literature search, coding, and
analysis.  The research ideas belong to the authors, and Claude was
used as an assistant under human supervision.  The authors reviewed
and edited all output and take full responsibility for the content of
the publication.


\clearpage
\section*{Online Resource 1: A worked example of the model generator and the complete list of benchmark instances}
\addcontentsline{toc}{section}{Online Resource 1}
\setcounter{section}{0}
\setcounter{equation}{0}
\setcounter{table}{0}
\renewcommand{\theequation}{S\arabic{equation}}
\renewcommand{\thetable}{S\arabic{table}}
\renewcommand{\thesection}{S\arabic{section}}

This supplement traces one of the smallest instances tabulated in the paper,
\code{2222\_3\_0} (Section~\ref{sec:all-instances}: size 5, optimum 2,
QUBO density 100\,\%), from the definitions of Section~2.3 of the paper
through to the literal files produced by the generator described in
Section~3 of the paper.

\section{Parameters and reduced space}
\label{sec:worked-example-space}

The instance is $H=H(2,2,2,2)$ (Section~2.3 of the paper), $d=3$, and
prune list $P=\{\overline{0}\}=\{0000\}$: the all-zero codeword is taken
as already belonging to the packing. Removing the open ball of radius
$d-1=2$ around it leaves the codewords at distance $\geq3$ from
$0000$, i.e.\ every weight-3 and weight-4 word:
$$
H\setminus B(0000,2) = \{0111,1011,1101,1110,1111\},
$$
which is exactly the reported size of 5. Since any two of these five
words differ in at most 2 positions (any two weight-3 words share a 1 in
at least one position, and every weight-3 word differs from $1111$ in
exactly one position), all $\binom{5}{2}=10$ pairs conflict: the reduced
conflict graph is complete, which is why the QUBO density for this
instance is 100\,\%.

\section{Generated files}

The three commands below (\code{"-o -"} without quote marks for output to standard output,
instead of a file) reproduce, respectively, the MILP, QUBO, and label
files for this instance:
\begin{verbatim}
./hampackmodgen.py -L -a 2222 -d 3 -p 0 -o -
./hampackmodgen.py -q -a 2222 -d 3 -p 0 -o -
./hampackmodgen.py --QUBO-LABELS -a 2222 -d 3 -p 0 -o -
\end{verbatim}

The MILP output is
{\small
\begin{verbatim}
\* Problem: None *\

Maximize
 obj: + objvalue

Subject To
 c0: x0000 + x0111 + x1011 + x1101 + x1110 + x1111 - objvalue = 0
 c1: x0111 + x1011 + x1101 + x1110 - x0000 >= 0
 c2: x0111 + x1011 <= 1
 c3: x0111 + x1101 <= 1
 c4: x0111 + x1110 <= 1
 c5: x0111 + x1111 <= 1
 c6: x0000 - x0111 >= 0
 c7: x1011 + x1101 <= 1
 c8: x1011 + x1110 <= 1
 c9: x1011 + x1111 <= 1
 c10: x0000 - x1011 >= 0
 c11: x1101 + x1110 <= 1
 c12: x1101 + x1111 <= 1
 c13: x0000 - x1101 >= 0
 c14: x1110 + x1111 <= 1
 c15: x0000 - x1110 >= 0
Bounds
 1 <= x0000 <= 1
 0 <= x0111 <= 1
 0 <= x1011 <= 1
 0 <= x1101 <= 1
 0 <= x1110 <= 1
 0 <= x1111 <= 1

Generals
 x0000
 x0111
 x1011
 x1101
 x1110
 x1111

End
\end{verbatim}
}
Two implementation details are worth pointing out, since they explain
why this file does not look like a direct transcription of
Eq.~(8) of the paper. First, the generator introduces an auxiliary
\code{objvalue} variable and its defining equality (\code{c0}) so that a
dual bound or the Komamiya--Joshi--Singleton bound
(Sections~2.2 and~3 of the paper)
can later be attached to
\code{objvalue} in \code{Bounds} without touching the objective row; this
example uses neither \code{-N}, \code{-k} nor \code{--dualbound}, so no
such bound appears. Second, the second constraint block of
Eq.~(8) (fixing the removed ball to $0$) is realized by
simply omitting those variables from the model altogether, rather than by
an explicit equality constraint -- equivalent, and smaller. With these
two exceptions, the correspondence with Eq.~(8) is direct:
\code{Bounds} fixes $x_{0000}=1$ (first block); constraints \code{c2--c5},
\code{c7--c9}, \code{c11}, \code{c12}, \code{c14} are the pairwise mutex
constraints (third block); and \code{c1}, \code{c6}, \code{c10},
\code{c13}, \code{c15} are the no-isolated-vertex cut
of Eq.~(7) of the paper (fourth block), one per codeword that has a
same-distance neighbor in the reduced space -- here, all five.

The QUBO output is
\begin{verbatim}
ICOORD,JCOORD,QVALUE
1,1,-1
1,2,1
1,3,1
1,4,1
1,5,1
2,2,-1
2,3,1
2,4,1
2,5,1
3,3,-1
3,4,1
3,5,1
4,4,-1
4,5,1
5,5,-1
\end{verbatim}
matching Eq.~(9) of the paper at the default penalty weight $\lambda=1$: a
$(-1)$ diagonal entry $Q_{aa}$ per codeword, and, since every pair among
the five conflicts (Section~\ref{sec:worked-example-space}), a $+1$
off-diagonal entry $Q_{ab}$ for
every one of the $\binom{5}{2}=10$ pairs -- hence the fully dense
upper-triangular block. The accompanying label file
\begin{verbatim}
1 0111
2 1011
3 1101
4 1110
5 1111
\end{verbatim}
recovers which codeword each QUBO index refers to, as described in
Section~3 of the paper.

\section{Cross-check}

Because the reduced conflict graph is complete, at most one of the five
codewords above can join $0000$ in a packing, so $N(H;3)=1+1=2$: e.g.\
$\{0000,1111\}$ has distance 4, and $\{0000,0111\}$ has distance 3, while
no two of $\{0111,1011,1101,1110,1111\}$ can be selected together. This
matches both the optimum reported for \code{2222\_3\_0} in
Section~\ref{sec:all-instances} and the value $N_2(4;d)=2$ tabulated
directly in \cite{Naszvadi2025}.

The remaining generator options not exercised here -- \code{-N},
\code{-k}, \code{--dualbound}, \code{-w}, \code{--nonzeros}, and the
\code{-C}/\code{--cliquer} output format -- are described in
Section~3 of the paper and, in full technical detail
(option semantics, exit statuses, file-naming conventions), in the
generator's own manpage and \code{README.org}, distributed alongside the
script in the research data.

\section{Complete list of benchmark instances with LP results}
\label{sec:all-instances}

Instances with MILP benchmark reference features. The optimal values come from the MILP solutions. The lines typed in boldface are the top-10 ones for which the largest number of MQLib algorithms failed. The italic lines are those for which SBM has not found the optimum. Instances marked with an asterisk are those which were addressed with DWave quantum hardware.

{\small
\begin{longtable}{lrrrrr}
\toprule
basename & size & optimum & LP runtime & LP ptime & QUBO density \\
         &      &         & (s)        & (s)      & (\%)         \\
\midrule
\endfirsthead
\toprule
basename & size & optimum & LP runtime & LP ptime & QUBO density \\
         &      &         & (s)        & (s)      & (\%)         \\
\midrule
\endhead
\midrule
\multicolumn{6}{r}{\textit{continued on next page}} \\
\midrule
\endfoot
\bottomrule
\endlastfoot
\multicolumn{6}{l}{\textbf{tiny}} \\
\midrule
\code{{222222\_4\_0}} & 22 & 4 & 0.01 & 0.00 & 82 \\
\code{{22222\_3\_0}} & 16 & 4 & 0.01 & 0.00 & 66 \\
\code{{2222\_3\_0}} & 5 & 2 & 0.00 & 0.00 & 100 \\
\code{{322222\_4\_0}} & 38 & 4 & 0.02 & 0.01 & 75 \\
\code{{32222\_3\_0}} & 27 & 6 & 0.01 & 0.01 & 55 \\
\code{{3222\_3\_0}} & 9 & 3 & 0.01 & 0.01 & 86 \\
\code{{332222\_4\_0}} & 65 & 6 & 0.06 & 0.05 & 67 \\
\code{{33222\_3\_0}} & 45 & 6 & 0.07 & 0.04 & 45 \\
\code{{3322\_3\_0}} & 16 & 4 & 0.01 & 0.00 & 73 \\
\code{{333222\_4\_0}} & 110 & 6 & 0.27 & 0.15 & 58 \\
\code{{33322\_3\_0}} & 74 & 9 & 0.28 & 0.27 & 36 \\
\code{{333322\_4\_0}} & 184 & 8 & 1.61 & 1.56 & 50 \\
\code{{*33333\_3\_0}} & 192 & 18 & 1.43 & 1.13 & 23 \\
\code{{42222222\_5\_0}} & 221 & 5 & 0.52 & 0.50 & 69 \\
\code{{42222222\_6\_0}} & 95 & 2 & 0.08 & 0.08 & 100 \\
\code{{4222222\_3\_0}} & 213 & 18 & 14.89 & 4.19 & 18 \\
\code{{4222222\_4\_0}} & 148 & 8 & 0.34 & 0.23 & 49 \\
\code{{4222222\_5\_0}} & 73 & 4 & 0.09 & 0.07 & 90 \\
\code{{422222\_3\_0}} & 94 & 9 & 0.79 & 0.38 & 30 \\
\code{{42222\_3\_0}} & 38 & 8 & 0.02 & 0.01 & 50 \\
\code{{4222\_3\_0}} & 13 & 4 & 0.01 & 0.01 & 80 \\
\code{{43222222\_6\_0}} & 168 & 3 & 0.28 & 0.25 & 96 \\
\code{{4322222\_4\_0}} & 242 & 8 & 8.12 & 1.01 & 42 \\
\code{{4322222\_5\_0}} & 127 & 4 & 0.16 & 0.15 & 83 \\
\code{{432222\_3\_0}} & 150 & 13 & 4.58 & 1.16 & 24 \\
\code{{4332222\_5\_0}} & 219 & 4 & 0.63 & 0.58 & 75 \\
\code{{433222\_3\_0}} & 237 & 18 & 172.21 & 32.57 & 19 \\
\code{{43322\_3\_0}} & 103 & 11 & 0.64 & 0.38 & 32 \\
\code{{433322\_4\_0}} & 258 & 8 & 22.26 & 0.96 & 46 \\
\textit{{43332\_3\_0}} & \textit{166} & \textit{14} & \textit{9.91} & \textit{4.07} & \textit{25} \\
\code{{433332\_5\_0}} & 208 & 4 & 1.40 & 0.99 & 82 \\
\code{{43333\_3\_0}} & 264 & 21 & 30.89 & 9.75 & 20 \\
\code{{4422222\_5\_0}} & 181 & 4 & 0.60 & 0.42 & 80 \\
\code{{442222\_3\_0}} & 206 & 16 & 0.83 & 0.82 & 21 \\
\code{{442222\_4\_0}} & 130 & 8 & 0.20 & 0.19 & 59 \\
\code{{44322\_3\_0}} & 143 & 12 & 0.46 & 0.46 & 28 \\
\textit{{44332\_3\_0}} & \textit{229} & \textit{18} & \textit{6.04} & \textit{6.03} & \textit{22} \\
\code{{443332\_5\_0}} & 300 & 4 & 16.92 & 16.73 & 79 \\
\code{{52222222\_5\_0}} & 285 & 6 & 1.02 & 0.99 & 67 \\
\code{{52222222\_6\_0}} & 124 & 2 & 0.17 & 0.16 & 100 \\
\code{{5222222\_3\_0}} & 270 & 20 & 68.02 & 8.29 & 16 \\
\code{{5222222\_4\_0}} & 190 & 8 & 0.68 & 0.67 & 47 \\
\code{{5222222\_5\_0}} & 95 & 4 & 0.09 & 0.08 & 88 \\
\textit{{522222\_3\_0}} & \textit{120} & \textit{11} & \textit{1.05} & \textit{1.05} & \textit{28} \\
\code{{53222222\_6\_0}} & 219 & 3 & 0.52 & 0.48 & 95 \\
\code{{5322222\_5\_0}} & 165 & 4 & 0.47 & 0.35 & 81 \\
\code{{532222\_3\_0}} & 191 & 16 & 1.15 & 1.15 & 22 \\
\code{{5332222\_5\_0}} & 284 & 5 & 1.19 & 1.16 & 74 \\
\textbf{{533222\_3\_0}} & \textbf{301} & \textbf{21} & \textbf{1202.81} & \textbf{11.32} & \textbf{17} \\
\code{{53322\_3\_0}} & 132 & 12 & 0.37 & 0.37 & 30 \\
\code{{53332\_3\_0}} & 212 & 15 & 241.31 & 124.49 & 23 \\
\code{{533332\_5\_0}} & 272 & 4 & 2.60 & 2.55 & 80 \\
\code{{5422222\_5\_0}} & 235 & 4 & 1.23 & 0.98 & 78 \\
\code{{62222222\_6\_0}} & 153 & 2 & 0.27 & 0.25 & 100 \\
\code{{6222222\_4\_0}} & 232 & 8 & 0.95 & 0.70 & 46 \\
\code{{622222\_3\_0}} & 146 & 12 & 1.10 & 1.03 & 26 \\
\code{{6322222\_5\_0}} & 203 & 4 & 0.71 & 0.54 & 80 \\
\code{{63332\_3\_0}} & 258 & 18 & 1.65 & 1.64 & 22 \\
\code{{6422222\_5\_0}} & 289 & 4 & 1.10 & 1.05 & 77 \\
\code{{72222222\_6\_0}} & 182 & 2 & 0.40 & 0.39 & 100 \\
\code{{7222222\_4\_0}} & 274 & 8 & 1.50 & 1.08 & 45 \\
\code{{*722222\_3\_0}} & 172 & 14 & 1.15 & 0.27 & 25 \\
\code{{722222\_4\_0}} & 102 & 4 & 0.17 & 0.13 & 69 \\
\code{{7322222\_5\_0}} & 241 & 4 & 0.92 & 0.79 & 79 \\
\code{{733222\_4\_0}} & 290 & 6 & 3.33 & 3.09 & 50 \\
\code{{822222\_3\_0}} & 198 & 16 & 0.57 & 0.34 & 25 \\
\multicolumn{6}{l}{\textbf{small}} \\
\midrule
\code{{42222222\_4\_0}} & 361 & 16 & 4.78 & 2.35 & 32 \\
\code{{43222222\_4\_0}} & 574 & 16 & 4084.29 & 7.43 & 27 \\
\code{{43222222\_5\_0}} & 369 & 6 & 3.51 & 3.45 & 61 \\
\code{{4332222\_4\_0}} & 392 & 12 & 140.90 & 2.67 & 35 \\
\code{{4333222\_5\_0}} & 374 & 6 & 5.16 & 3.43 & 68 \\
\textit{\textbf{{433322\_3\_0}}} & \textit{\textbf{371}} & \textit{\textbf{27}} & \textit{\textbf{2851.54}} & \textit{\textbf{350.40}} & \textit{\textbf{15}} \\
\code{{433332\_4\_0}} & 424 & 12 & 143.09 & 66.68 & 38 \\
\code{{433333\_4\_0}} & 688 & 18 & 1474.12 & 183.78 & 32 \\
\code{{433333\_5\_0}} & 368 & 4 & 27.83 & 27.75 & 73 \\
\textbf{{4422222\_3\_0}} & \textbf{451} & \textbf{32} & \textbf{23.99} & \textbf{23.97} & \textbf{12} \\
\code{{4422222\_4\_0}} & 336 & 9 & 54.39 & 6.48 & 38 \\
\textbf{{4432222\_4\_0}} & \textbf{542} & \textbf{13} & \textbf{2540.64} & \textbf{107.71} & \textbf{32} \\
\code{{4432222\_5\_0}} & 311 & 5 & 1.76 & 1.72 & 72 \\
\code{{443222\_3\_0}} & 324 & 24 & 1.92 & 0.00 & 16 \\
\textbf{{*443322\_3\_0}} & \textbf{505} & \textbf{36} & \textbf{283.11} & \textbf{283.08} & \textbf{13} \\
\code{{443322\_4\_0}} & 361 & 10 & 52.99 & 14.76 & 42 \\
\textit{{443332\_4\_0}} & \textit{590} & \textit{14} & \textit{1245.96} & \textit{563.76} & \textit{35} \\
\code{{44333\_3\_0}} & 362 & 27 & 18.48 & 18.46 & 17 \\
\textbf{{444222\_3\_0}} & \textbf{442} & \textbf{32} & \textbf{39.10} & \textbf{39.08} & \textbf{14} \\
\code{{444322\_4\_0}} & 504 & 12 & 33.15 & 33.08 & 38 \\
\code{{52222222\_4\_0}} & 460 & 16 & 9.32 & 4.38 & 31 \\
\code{{53222222\_4\_0}} & 730 & 16 & 9294.02 & 15.95 & 25 \\
\code{{53222222\_5\_0}} & 475 & 7 & 8.02 & 7.68 & 59 \\
\code{{5322222\_4\_0}} & 310 & 8 & 16.89 & 2.00 & 40 \\
\code{{5332222\_4\_0}} & 501 & 12 & 366.97 & 1.21 & 33 \\
\code{{533322\_4\_0}} & 332 & 9 & 32.81 & 5.32 & 44 \\
\code{{533332\_4\_0}} & 544 & 12 & 661.95 & 53.34 & 36 \\
\code{{53333\_3\_0}} & 336 & 21 & 21191.28 & 101.56 & 18 \\
\code{{5422222\_4\_0}} & 430 & 9 & 241.74 & 191.48 & 37 \\
\code{{5432222\_4\_0}} & 692 & 13 & 39771.30 & 65.49 & 30 \\
\code{{5432222\_5\_0}} & 403 & 6 & 4.14 & 4.07 & 70 \\
\textbf{{543322\_3\_0}} & \textbf{639} & \textbf{36} & \textbf{105.50} & \textbf{105.46} & \textbf{11} \\
\code{{543322\_4\_0}} & 464 & 11 & 196.96 & 33.14 & 40 \\
\code{{5522222\_4\_0}} & 550 & 11 & 171.81 & 55.70 & 35 \\
\code{{5522222\_5\_0}} & 305 & 4 & 1.27 & 1.21 & 76 \\
\code{{553322\_4\_0}} & 596 & 12 & 45.96 & 45.86 & 38 \\
\code{{62222222\_4\_0}} & 559 & 16 & 17.47 & 8.67 & 30 \\
\code{{62222222\_5\_0}} & 349 & 6 & 1.73 & 1.66 & 65 \\
\code{{6222222\_3\_0}} & 327 & 24 & 76.21 & 0.00 & 15 \\
\textit{\textbf{{6322222\_3\_0}}} & \textit{\textbf{508}} & \textit{\textbf{32}} & \textit{\textbf{88.28}} & \textit{\textbf{88.25}} & \textit{\textbf{12}} \\
\code{{6322222\_4\_0}} & 378 & 8 & 30.81 & 3.28 & 39 \\
\code{{6332222\_4\_0}} & 610 & 12 & 540.97 & 1.91 & 32 \\
\code{{6332222\_5\_0}} & 349 & 6 & 2.08 & 2.03 & 72 \\
\textbf{{633222\_3\_0}} & \textbf{365} & \textbf{24} & \textbf{4.07} & \textbf{4.06} & \textbf{16} \\
\textbf{{633322\_3\_0}} & \textbf{569} & \textbf{36} & \textbf{54.30} & \textbf{54.26} & \textbf{12} \\
\code{{633332\_4\_0}} & 664 & 12 & 952.54 & 269.09 & 35 \\
\code{{63333\_3\_0}} & 408 & 24 & 8466.46 & 889.74 & 17 \\
\code{{6422222\_4\_0}} & 524 & 9 & 358.58 & 349.44 & 35 \\
\textit{{6522222\_4\_0}} & \textit{670} & \textit{11} & \textit{429.16} & \textit{80.17} & \textit{33} \\
\code{{6522222\_5\_0}} & 375 & 4 & 2.18 & 2.08 & 75 \\
\textbf{{6665\_3\_0}} & \textbf{925} & \textbf{30} & \textbf{118.58} & \textbf{118.47} & \textbf{15} \\
\code{{72222222\_4\_0}} & 658 & 16 & 28.82 & 28.72 & 29 \\
\code{{72222222\_5\_0}} & 413 & 7 & 2.54 & 2.47 & 64 \\
\code{{7222222\_3\_0}} & 384 & 28 & 70.08 & 4.53 & 15 \\
\code{{7322222\_4\_0}} & 446 & 8 & 42.55 & 40.89 & 38 \\
\textbf{{73333\_3\_0}} & \textbf{480} & \textbf{27} & \textbf{13.49} & \textbf{13.46} & \textbf{16} \\
\code{{7422222\_4\_0}} & 618 & 9 & 253.82 & 195.77 & 34 \\
\code{{7422222\_5\_0}} & 343 & 4 & 1.59 & 1.53 & 76 \\
\textbf{{7666\_3\_0}} & \textbf{1325} & \textbf{36} & \textbf{4206.21} & \textbf{4206.02} & \textbf{12} \\
\textbf{{82222222\_3\_0}} & \textbf{939} & \textbf{64} & \textbf{128.09} & \textbf{128.03} & \textbf{8} \\
\textbf{{8222222\_3\_0}} & \textbf{441} & \textbf{32} & \textbf{4.60} & \textbf{4.58} & \textbf{14} \\
\code{{8222222\_4\_0}} & 316 & 8 & 1.90 & 1.46 & 44 \\
\code{{8322222\_4\_0}} & 514 & 8 & 40.72 & 9.02 & 37 \\
\code{{9222222\_4\_0}} & 358 & 8 & 2.37 & 1.91 & 44 \\
\code{{9322222\_4\_0}} & 582 & 8 & 62.20 & 30.28 & 37 \\
\code{{933322\_4\_0}} & 628 & 9 & 94.25 & 33.44 & 40 \\
\code{{b43322\_4\_0}} & 1082 & 11 & 1037.71 & 300.13 & 35 \\
\end{longtable}
}

\end{document}